\documentclass[12pt]{article}

\usepackage{amssymb}
\usepackage{amsmath}

\usepackage{multirow}

\usepackage{graphicx}

\newcommand{\qed}{\hfill \rule{2.3mm}{2.3mm}}
\newcommand{\fr}[2]{\frac{\displaystyle #1}{\displaystyle #2}}
\newcommand{\df}[2]{\frac{\displaystyle d#1}{\displaystyle d#2}}

\newcommand{\Real}{\mathrm{Re}}
\newcommand{\Imag}{\mathrm{Im}}

\newtheorem{theorem}{Theorem}[section]

\newtheorem{proposition}[theorem]{Proposition}
\newtheorem{remark}[theorem]{Remark}

\begin{document}
\title{Periodic orbits in the Ott-Antonsen manifold}
\author{O. E. Omel'chenko}
%\affiliation{Institute of Physics and Astronomy, University of Potsdam,
%Karl-Liebknecht-Str. 24/25,
%14476 Potsdam,
%Germany}
%

\maketitle

\begin{abstract}
In their seminal paper [Chaos 18, 037113 (2008)], E. Ott and T. M. Antonsen showed
that large groups of phase oscillators driven by a certain type of common force
display low dimensional long-term dynamics,
which is described by a small number of ordinary differential equations.
This fact was later used as a simplifying reduction technique in many studies of synchronization phenomena
occurring in networks of coupled oscillators and in neural networks.
Most of these studies focused mainly on partially synchronized states
corresponding to the equilibrium-type dynamics
in the so called Ott-Antonsen manifold.
Going beyond this paradigm, here we propose a universal approach for the efficient analysis
of partially synchronized states with non-equilibrium periodic collective dynamics.
Our method is based on the observation
that the Poincar{\'e} map of the complex Riccati equation,
which describes the dynamics in the Ott-Antonsen manifold,
coincides with the well-known M{\"o}bius transformation.
To illustrate the possibilities of our method, we use it to calculate
a complete bifurcation diagram of travelling chimera states
in a ring network of phase oscillators with asymmetric nonlocal coupling.
\end{abstract}

\section{Introduction}

Synchronization of rhythmic processes is a fundamental dynamical mechanism
that underlies the functioning of many natural
and man-made systems~\cite{Book:PikovskyRK,Book:Kuramoto,Book:Winfree,Gla2001}.
Circadian clocks~\cite{YamIMOYKO2013} and metachronal waves in cilia carpets~\cite{ElgG2013,SolF2022},
Josephson junction arrays~\cite{WieCS1996} and power grids~\cite{RohSTW2012,MotMAN2013}
are just a few examples of this kind.
Other situations are also known where synchronization can occur, but is undesirable.
These include different neurological disorders such as schizophrenia, epilepsy,
Alzheimer's and Parkinson's disease~\cite{UhlS2006,LehBHKRSW2009}.
The variety of the above examples was the motivation
for the development of mathematical theory of synchronization~\cite{AreD-GKMZ2008,Book:BoccalettiPGA},
culminating in the method of master stability function~\cite{PecC1998}
and numerous case studies for the paradigmatic Kuramoto model~\cite{Str2000,AceBVRS2005,PikR2015,RodPJK2016}.
Initially, the main focus of research was on full synchronization~\cite{DoeCB2013},
when all components of a system behave identically,
but later more complex forms of synchronous collective dynamics were identified and studied,
including clustered synchronization~\cite{HanMM1993,SorPHMR2016}, generalized synchronization~\cite{RulSTA1996},
phase synchronization of chaos~\cite{RosPK1996} and self-organized quasiperiodicity~\cite{RosP2007}.

In heterogeneous systems consisting of many non-identical oscillators,
synchronization usually occurs as a dynamical aggregation process
controlled by the intensity of the interaction between these oscillators.
The course of the process depends on the properties of the oscillators
and the nature of their interaction~\cite{Paz2005,OmeW2012,Lai2009,Gom-GMA2007,OmeWL2014}.
Roughly speaking, for each such system there is a critical coupling strength,
above which the asynchronous state becomes unstable,
while one or more frequency-synchronized clusters are formed.
As the coupling strength increases, the clusters increase in size and merge together,
so that a fully synchronized state is eventually achieved.
The simplest and most detailed description of this phenomenon
can be given using the Kuramoto model~\cite{Str2000},
where the state of each oscillator is represented by one scalar variable, its phase,
and the oscillators are all-to-all coupled.
More specifically, if $\omega_j\in\mathbb{R}$ denotes the natural frequency of the $j$th oscillator
and $K > 0$ is the coupling strength, then the phases of the oscillators evolve according to
\begin{equation}
\df{\theta_j}{t} = \omega_j - \fr{K}{N} \sum\limits_{k=1}^N \sin( \theta_j - \theta_k ),\qquad j=1,\dots,N.
\label{Eq:Kuramoto}
\end{equation}
A remarkable feature of model~(\ref{Eq:Kuramoto}) is that
in the thermodynamic limit $N\to\infty$ its dynamics can be carefully investigated
using the McKean-Vlasov equation (also called the continuity equation)
for a distribution $\rho(\theta,\omega,t)$ that yields the probability to find
an oscillator $\theta_j(t)\approx\theta$ with the natural frequency $\omega_j\approx\omega$ at time $t$.
Moreover, if the natural frequencies $\omega_j$ are drawn from a Lorentzian distribution,
then the long-term dynamics of $\rho(\theta,\omega,t)$ converges to a low-dimensional manifold
parameterized by the limit value (as $N\to\infty$) of the global order parameter
\begin{equation}
z(t) = \fr{1}{N} \sum\limits_{k=1}^N e^{i \theta_k(t)}.
\label{Def:OP}
\end{equation}
This manifold was first described by Ott and Antonsen in~\cite{OttA2008,OttA2009},
and since then, it has been used to study various types of collective dynamics
in systems of phase oscillators~\cite{Lai2009,Lai2009a}, Winfree oscillators~\cite{PazM2014},
theta neurons~\cite{LukBS2013,OmeL2022}, and quadratic integrate-and-fire neurons~\cite{MonPR2015,ByrAC2019}.
It should be noted that so far the full power of the Ott-Antonsen method
has been demonstrated mainly for statistically stationary states.
For example, in the Kuramoto model~(\ref{Eq:Kuramoto})
such are asynchronous and partially synchronized states
with nearly constant magnitudes $|z(t)|$ of the order parameter~(\ref{Def:OP}).
On the other hand, there are a number of cases where the Ott-Antonsen method
allowed to detect more complex collective dynamics of oscillators and neurons,
which has so far been studied only superficially.
In this regard, we can mention non-stationary partially synchronized states with periodically oscillating magnitudes $|z(t)|$,
which occur in the model~(\ref{Eq:Kuramoto}) with a bimodal distribution of natural frequencies~$\omega_j$~\cite{MarBSOSA2009},
as well as other examples of periodically modulated collective dynamics, which were briefly reported in~\cite{Lai2009,LukBS2013}.

Motivated by the latter examples, in this paper we want to propose a mathematical approach
for the efficient analysis of periodic dynamics in the Ott-Antonsen manifold.
For this, in Section~\ref{Sec:Models} we introduce two auxiliary models:
(i) a system of phase oscillators driven by a common periodic force,
and (ii) a system of theta neurons driven by a common periodic input current.
For both models we write complex differential equations
that determine the corresponding collective dynamics in the Ott-Antonsen manifold.
In Section~\ref{Sec:Riccati} we consider these equations
and show that their Poincar{\'e} maps coincide with the well-known M{\"o}bius transformation.
Next, in Section~\ref{Sec:Moebius} we explain how this fact can be used
for fast calculation of periodic orbits in the Ott-Antonsen manifold.
A practical application of the proposed semi-analytical method
is described in Section~\ref{Sec:Chimera}.
There, we consider a nonlinear integro-differential equation of the Ott-Antonsen type,
which describes the long-term dynamics of a ring network
of nonlocally coupled phase oscillators with Lorentzian-distributed natural frequencies.
We focus on the travelling wave solutions of this equation,
which represent so called travelling chimera states~\cite{Ome2020},
and show how these solutions can be quickly calculated
using a kind of self-consistency argument.
Finally, in Section~\ref{Sec:Discussion} we discuss other problems,
which can be considered by the method of this paper.

\section{Two models and two Ott-Antonsen equations}
\label{Sec:Models}

The simplest and therefore the most popular mathematical models
used in the study of synchronization phenomena
include various types of networks consisting of phase oscillators and theta-neurons.
Their collective dynamics is often investigated by means of the self-consistency method.
Roughly speaking, one assumes that oscillators or neurons are influenced by a given external force
and calculates the corresponding dynamics of each network's node.
From this, the effective force of interaction between nodes is estimated
and the obtained value is compared with the initially assumed value of the force.
The resulting match relation is called the self-consistency equation
and its analysis is often much easier than the analysis of the original network dynamics.
Below, we describe two auxiliary models related to the application of the self-consistency method
to networks of phase oscillators and networks of theta-neurons.
%Note that since we are going to focus on periodically modulated collective dynamics,
%we consider two models with periodic driving forces.

{\it Phase oscillators.} Using definition~(\ref{Def:OP}),
the Kuramoto model~(\ref{Eq:Kuramoto}) can be rewritten in the form
$$
\df{\theta_j}{t} = \omega_j + \Imag\left( K z(t) e^{-i \theta_j} \right),\qquad j=1,\dots,N,
$$
as if each oscillator is driven by the global order parameter~$z(t)$ multiplied by~$K$.
Generalizing this setting, we can write another model:
a population of phase oscillators
driven by an arbitrary complex-valued force $W(t)$
\begin{equation}
\df{\theta_j}{t} = \omega_j + \Imag\left( W(t) e^{-i \theta_j} \right),\qquad j=1,\dots,N,
\label{Eq:Oscillators}
\end{equation}
where $\theta_j(t)\in\mathbb{R}$ is still the phase of the $j$th oscillator
and $\omega_j\in\mathbb{R}$ is its natural frequency.
If the natural frequencies~$\omega_j$ are drawn randomly and independently
from a Lorentzian distribution
$$
g(\omega) = \fr{\gamma}{\pi}\fr{1}{(\omega - \omega_0)^2 + \gamma^2}\quad\mbox{with}\quad \omega_0\in\mathbb{R}\quad\mbox{and}\quad\gamma > 0,
$$
then using the Ott-Antonsen method~\cite{OttA2008,OttA2009}
it can be shown~\cite[Sec.~3.1]{BicGLM2020} that in the thermodynamics limit $N\to\infty$
the long-term dynamics of system~(\ref{Eq:Oscillators}) is completely described
by a scalar complex equation
\begin{equation}
\df{z}{t} = ( - \gamma + i \omega_0 ) z + \fr{1}{2} W(t) - \fr{1}{2} \overline{W(t)} z^2.
\label{Eq:OA:Oscillators}
\end{equation}
We call this equation the Ott-Antonsen equation corresponding to model~(\ref{Eq:Oscillators}).

{\it Theta neurons.} The theta neuron is an excitable dynamical unit
that is described by the normal form
of a saddle-node-on-an-invariant-circle (SNIC) bifurcation~\cite{ErmK1986,LukBS2013}.
The external driving usually acts on each neuron in the form of a real input current $J(t)$ so that
\begin{equation}
\df{\theta_j}{t} = 1 - \cos\theta_j + ( 1 + \cos\theta_j ) ( \eta_j + J(t) ),\qquad j=1,\dots,N,
\label{Eq:Neurons}
\end{equation}
where $\eta_j\in\mathbb{R}$ is the excitability parameter of the $j$th neuron.
In the case when $\eta_j$ are chosen from a Lorentzian distribution
$$
h(\eta) = \fr{\gamma}{\pi}\fr{1}{(\eta - \eta_0)^2 + \gamma^2}\quad\mbox{with}\quad \eta_0\in\mathbb{R}\quad\mbox{and}\quad \gamma > 0,
$$
one can also apply the Ott-Antonsen theory and obtain the mean-field equation~\cite[Sec.~3.1]{BicGLM2020}
\begin{equation}
\df{z}{t} = \fr{ ( - \gamma + i \eta_0 + i J(t) ) (1 + z)^2 - i (1 - z)^2 }{2}.
\label{Eq:OA:Neurons}
\end{equation}
By analogy with~(\ref{Eq:OA:Oscillators}), we call Eq.~(\ref{Eq:OA:Neurons})
the Ott-Antonsen equation corresponding to model~(\ref{Eq:Neurons}).

\section{Complex Riccati equation}
\label{Sec:Riccati}

It is easy to see that both Eq.~(\ref{Eq:OA:Oscillators}) and Eq.~(\ref{Eq:OA:Neurons})
are particular cases of the more general {\it complex Riccati equation}
\begin{equation}
\df{z}{t} = c_0(t) + c_1(t) z + c_2(t) z^2.
\label{Eq:Riccati}
\end{equation}
Indeed, Eq.~(\ref{Eq:Riccati}) coincides with Eq.~(\ref{Eq:OA:Oscillators}), if we assume
\begin{equation}
c_0(t) = \fr{1}{2} W(t),\qquad c_1(t) = - \gamma + i \omega_0,\qquad c_2(t) = - \fr{1}{2} \overline{W(t)}.
\label{c012:Oscillators}
\end{equation}
Similarly, Eq.~(\ref{Eq:OA:Neurons}) is obtained from Eq.~(\ref{Eq:Riccati}) in the case
\begin{equation}
c_0(t) = c_2(t) = \fr{1}{2} ( - \gamma + i ( \eta_0 + J(t) - 1 ) ),\qquad
c_1(t) =  - \gamma + i ( \eta_0 + J(t) + 1 ).
\label{c012:Neurons}
\end{equation}

Recall that in the Ott-Antonsen method, the function $z(t)$,
which solves Eq.~(\ref{Eq:OA:Oscillators}) or Eq.~(\ref{Eq:OA:Neurons}),
determines the limit value (as $N\to\infty$) of the global order parameter~(\ref{Def:OP}),
so by definition it must satisfy the inequality $|z(t)| \le 1$ for all $t\ge 0$.
Therefore, in the rest of the section we consider only such complex Riccati equations,
which guarantee this property. For a more accurate statement,
let us denote by $\mathbb{D} = \{ z\in\mathbb{C}\::\: |z| < 1 \}$ the open unit disc in the complex plane,
by $\overline{\mathbb{D}} = \{ z\in\mathbb{C}\::\: |z| \le 1 \}$ the closure of $\mathbb{D}$,
and by $\mathbb{S} = \{ z\in\mathbb{C}\::\: |z| = 1 \}$ the boundary of $\mathbb{D}$.
Then the next proposition provides a sufficient condition
for Eq.~(\ref{Eq:Riccati}) to be consistent with the Ott-Antonsen method.

\begin{proposition}
Suppose
$$
\Real( \overline{c_0(t)} z + c_1(t) + c_2(t) z ) \le 0\quad\mbox{for all}\quad z\in\mathbb{S}\quad\mbox{and}\quad t\ge 0,
$$
then the closed unit disc $\overline{\mathbb{D}}$ is an invariant set of Eq.~(\ref{Eq:Riccati}). In other words, if $z(0)\in\overline{\mathbb{D}}$,
then the corresponding solution $z(t)$ of Eq.~(\ref{Eq:Riccati}) lies in $\overline{\mathbb{D}}$ for all $t > 0$.
\label{Proposition:D}
\end{proposition}

{\bf Proof:} Let $z(t)$ be a solution of Eq.~(\ref{Eq:Riccati}), then simple calculation yield
\begin{equation}
\df{|z|^2}{t} = \overline{z} \df{z}{t} + z \df{\overline{z}}{t} = 2 \Real\left( \overline{c_0(t)} z + ( c_1(t) + c_2(t) z ) |z|^2 \right).
\label{Eq:Abs}
\end{equation}
For $|z| = 1$, this equation implies
$$
\df{|z|^2}{t} = 2 \Real\left( \overline{c_0(t)} z + c_1(t) + c_2(t) z \right) \le 0.
$$
Hence, if $|z(0)| \le 1$, then $|z(t)|$ cannot grow above one,
and therefore $|z(t)| \le 1$ for all $t\ge 0$.~\qed

\begin{remark}
The requirements of Proposition~\ref{Proposition:D} are fulfilled for Eqs.~(\ref{Eq:OA:Oscillators}) and~(\ref{Eq:OA:Neurons}), if $\gamma\ge 0$.
\end{remark}

Now we focus on the periodic complex Riccati equation, i.e. Eq.~(\ref{Eq:Riccati})
where $c_0(t)$, $c_1(t)$ and $c_2(t)$ are $2\pi$-periodic continuous complex-valued functions.
Our goal is to analyze all possible $2\pi$-periodic solutions of this equation
and determine the conditions that ensure the existence of a periodic solution
that lies entirely in the unit disc $\mathbb{D}$.

Let $U(t,z_0)$ denote the solution of the initial value problem for Eq.~(\ref{Eq:Riccati}) with the initial condition~$z(0) = z_0$.
Then the mapping $z_0\in\mathbb{C}\mapsto U(2\pi,z_0)\in\mathbb{C}$ is called the Poincar{\'e} map of Eq.~(\ref{Eq:Riccati}).
For the periodic complex Riccati equation, it is known~\cite{Cam1997,Wil2008}
that its Poincar{\'e} map is a one-to-one map of the extended complex plane
$\hat{\mathbb{C}} = \mathbb{C}\cup\{\infty\}$ (also called the Riemann sphere) onto itself,
which corresponds to the M{\"o}bius transformation
\begin{equation}
\mathcal{M}(z) = \fr{a z + b}{c z +d}
\label{Def:M}
\end{equation}
with complex coefficients $a$, $b$, $c$ and~$d$ such that $a d - b c \ne 0$.
This correspondence provides a useful mathematical tool for analyzing the periodic solutions of Eq.~(\ref{Eq:Riccati}).
Indeed, every periodic solution of Eq.~(\ref{Eq:Riccati}) corresponds to a fixed point of the transformation~$\mathcal{M}(z)$.
Since the M{\"o}bius transformation in general has two different fixed points
(or one fixed point of multiplicity two in the degenerate case),
the same can be said about the number of periodic solutions of Eq.~(\ref{Eq:Riccati}).
More detailed information about the position and stability of these fixed points
can be obtained from the geometric properties of the M{\"o}bius transformation~\cite{Book:Needham}.
For this, one needs to consider the behaviour of map trajectories,
i.e. complex sequences $z_{n+1} = \mathcal{M}(z_n)$, $n = 0,1,2,\dots$,
with different initial conditions~$z_0\in\mathbb{C}$.
Note that every M{\"o}bius transformation is invertible.
For instance, the inverse of~(\ref{Def:M}) reads
$$
\mathcal{M}^{-1}(z) = \fr{d z - b}{-c z + a}.
$$
Therefore, the above map trajectory can be extended
not only for increasing indices~$n$ but also for decreasing ones,
namely $z_{n-1} = \mathcal{M}^{-1}(z_n)$, $n = 0,-1,-2,\dots$.

\begin{figure}[h]
\centering
\includegraphics[width=0.24\textwidth,angle=270]{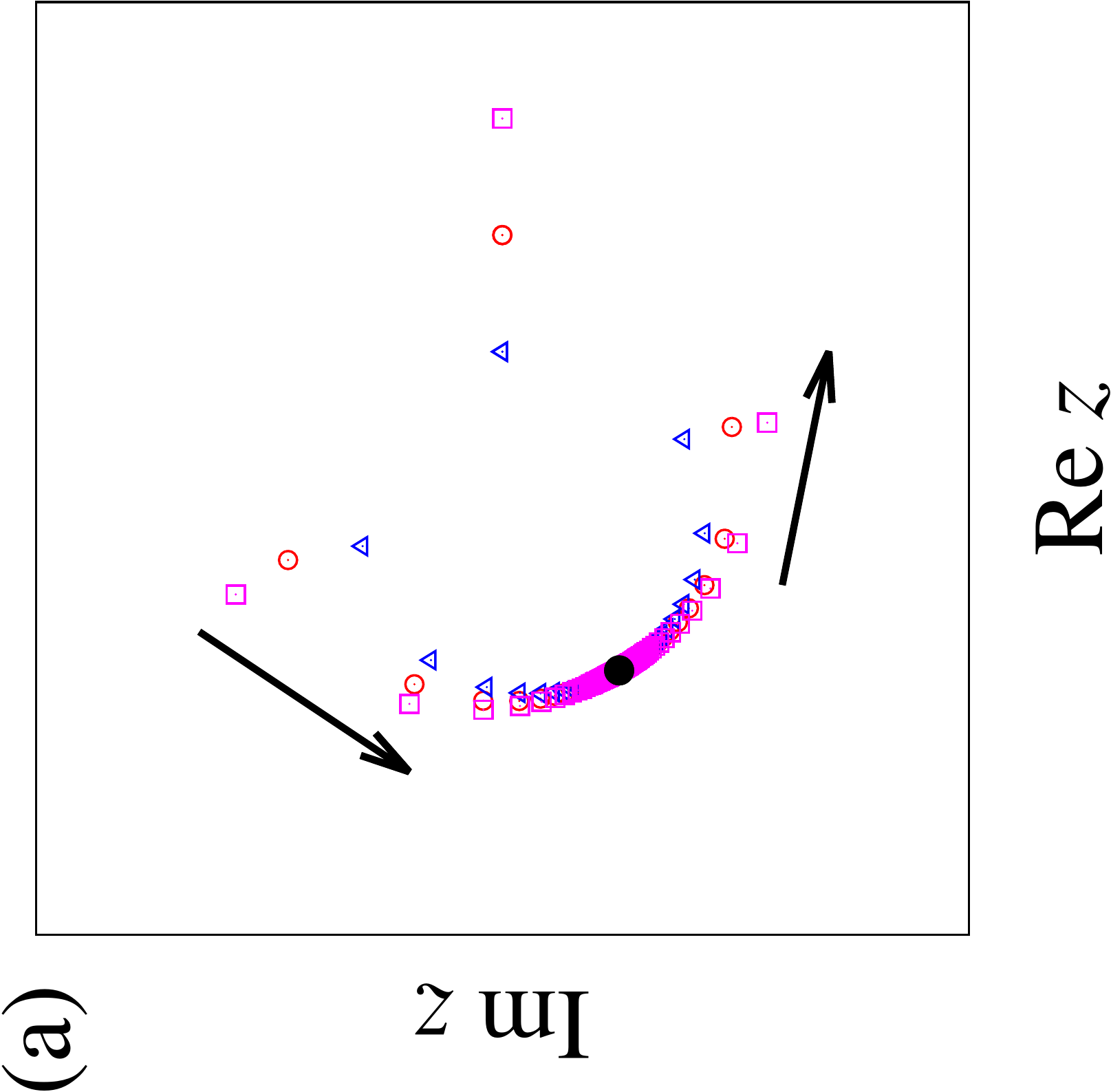}\hspace{0.013\textwidth}%
\includegraphics[width=0.24\textwidth,angle=270]{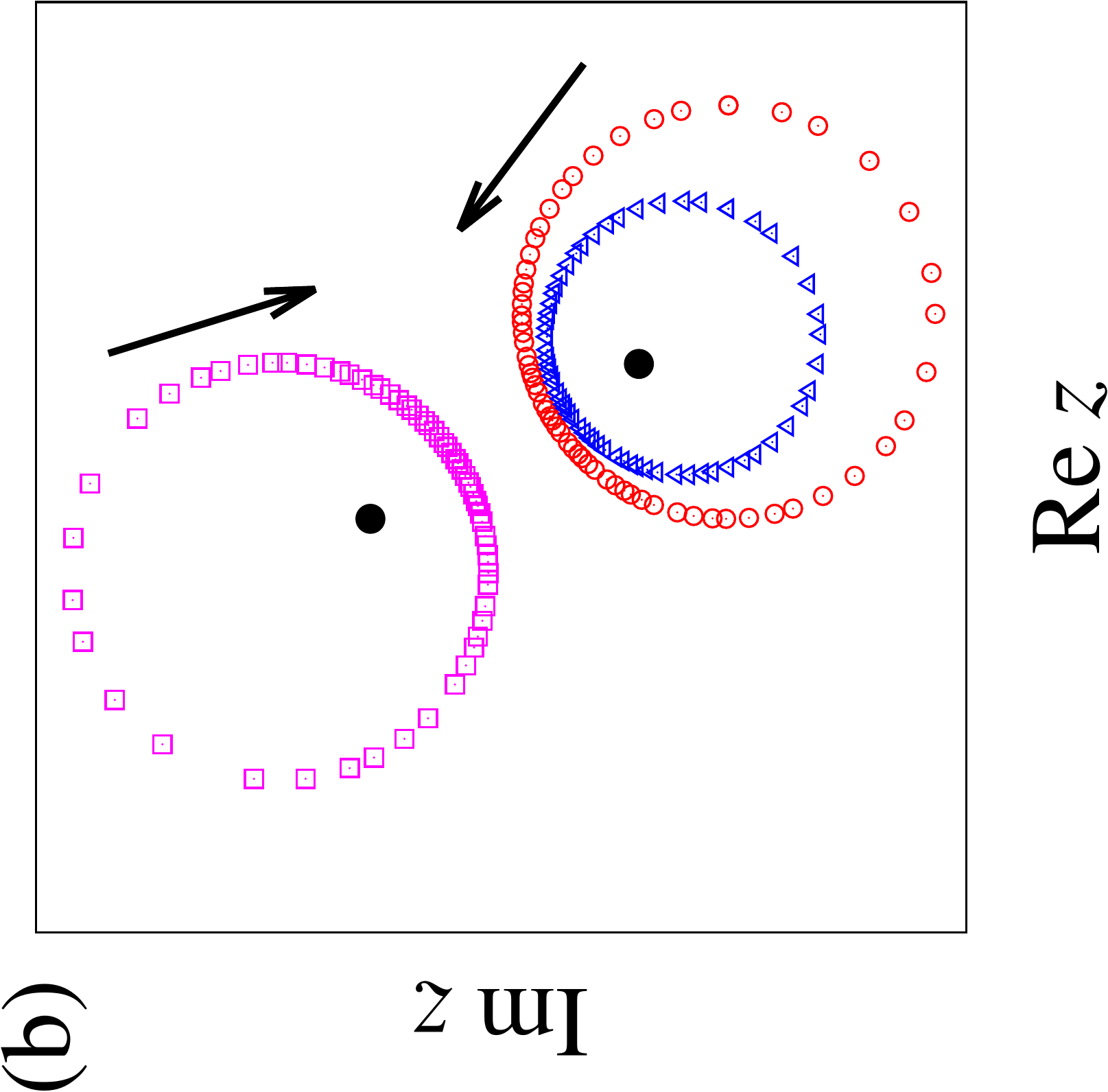}\hspace{0.013\textwidth}%
\includegraphics[width=0.24\textwidth,angle=270]{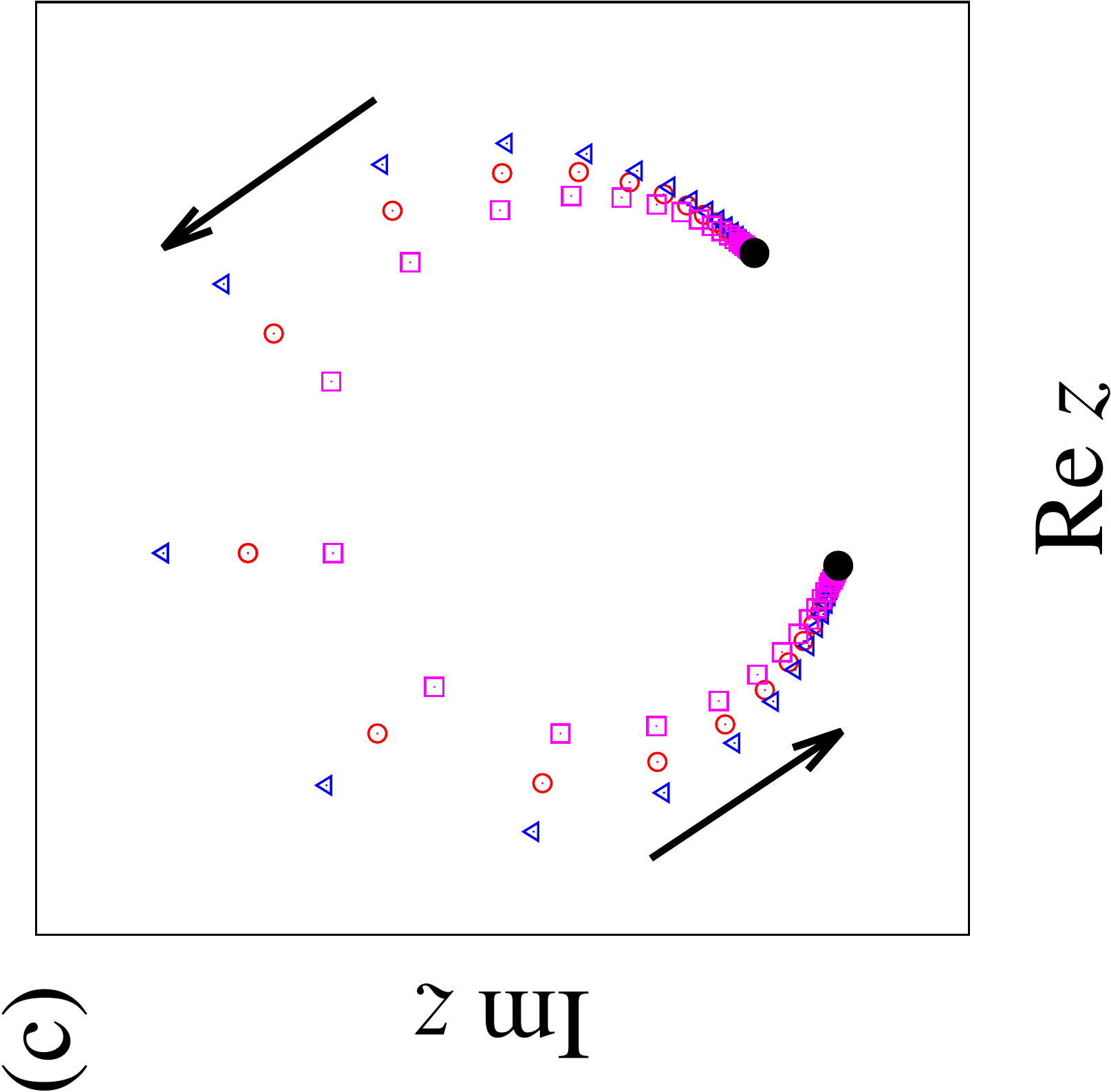}\hspace{0.013\textwidth}%
\includegraphics[width=0.24\textwidth,angle=270]{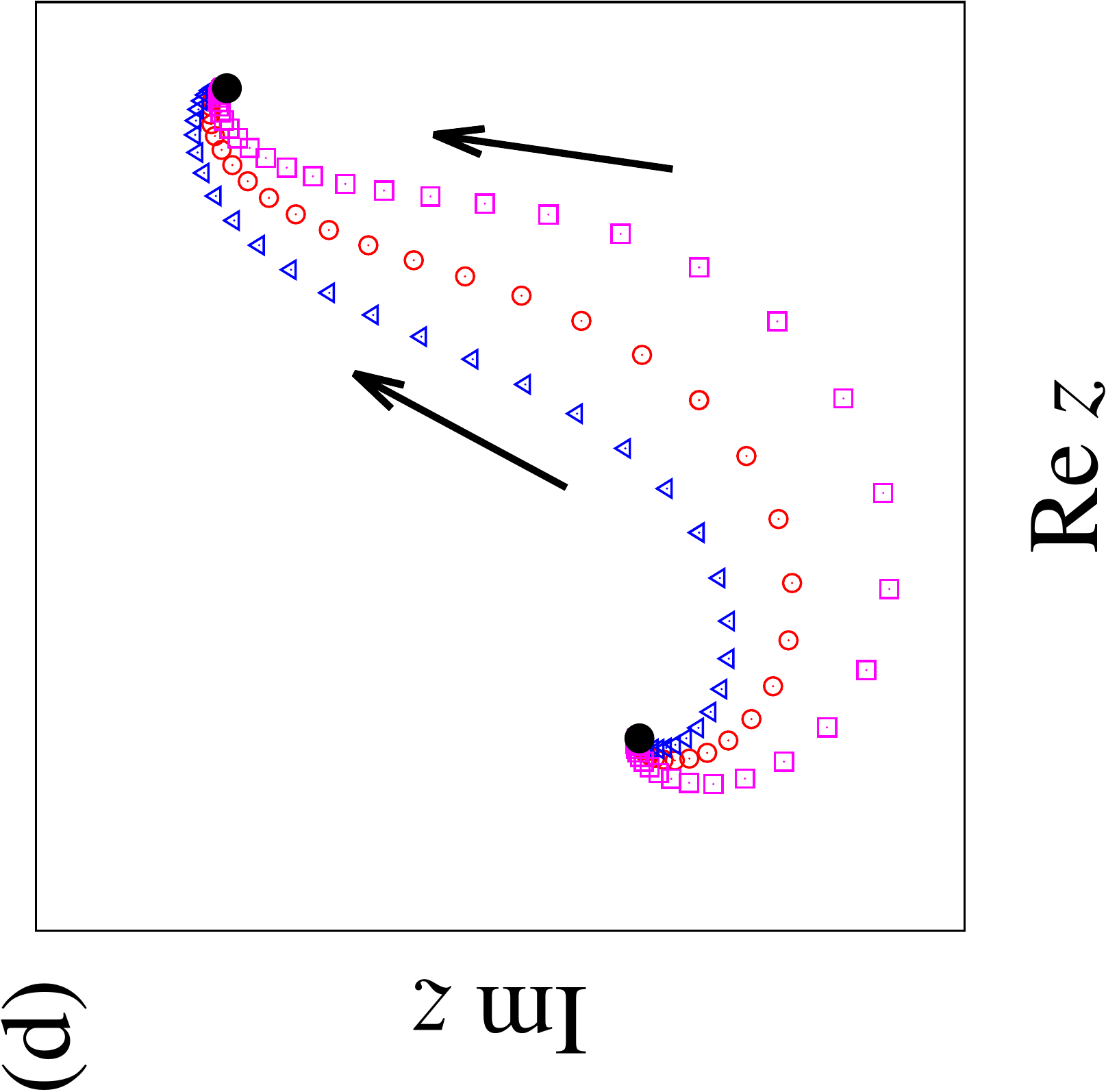}
\caption{Four main types of M{\"o}bius transformations:
(a) parabolic, (b) elliptic, (c) hyperbolic, and (d) loxodromic.
Empty circles, squares and triangles show three typical map trajectories.
Arrows indicate the direction of discrete time~$n$.
Filled circles show the position of fixed points.}
\label{Fig:Moebius}
\end{figure}

Qualitative difference in the behaviour of map trajectories
allows one to identify four main types of M{\"o}bius transformations:
parabolic, elliptic, hyperbolic and loxodromic, see Fig.~\ref{Fig:Moebius}.
Parabolic transform is the only type of M{\"o}bius transformation
with one degenerate fixed point of multiplicity two.
In this case, every map trajectory converges to this fixed point for both $n\to-\infty$ and $n\to+\infty$, e.g. Fig.~\ref{Fig:Moebius}(a).
In contrast, all non-parabolic transforms have two different fixed points.
More specifically, for an elliptic transform each map trajectory lies on a circle around one of the fixed points, e.g. Fig.~\ref{Fig:Moebius}(b).
In the resonant case, the trajectory visits only a finite number of points of the circle,
otherwise it fills the circle densely. For hyperbolic and loxodromic transforms,
each map trajectory lies on a smooth curve that connects two fixed points,
see Fig.~\ref{Fig:Moebius}(c) and (d) respectively.
The trajectory converges to one of the fixed points for $n\to+\infty$ and to the other for $n\to-\infty$.
The main difference between hyperbolic and loxodromic cases originates from the fact
that in the former case the curves on which the different trajectories lie are circular arcs,
while in the latter case they are logarithmic spirals.
%%%%%%%%%%%%%%%%%%%%%%%%%%%%%%%%%%%%%%%
%%%%%%%%%%%%%%%%%%%%%%%%%%%%%%%%%%%%%%%

The location of the map trajectories in each of the above cases
determines the stability of the fixed points of the corresponding M{\"o}bius transformation.
For example, we can see that the degenerate fixed point of a parabolic transform
is always unstable in the sense of Lyapunov. On the other hand, both fixed points
of an elliptic transform are stable in the sense of Lyapunov, but not asymptotically stable.
Finally, every hyperbolic or loxodromic transform has one asymptotically stable and one unstable fixed point.
These facts are important for our further consideration,
because they allow us to characterize the stability of periodic solutions of Eq.~(\ref{Eq:Riccati}).

\begin{proposition}
Suppose
\begin{equation}
\Real( \overline{c_0(t)} z + c_1(t) + c_2(t) z ) < 0\quad\mbox{for all}\quad z\in\mathbb{S}\quad\mbox{and}\quad 0\le t \le 2\pi,
\label{Ineq:c}
\end{equation}
then the Poincar{\'e} map of Eq.~(\ref{Eq:Riccati}) is described by a hyperbolic or loxodromic M{\"o}bius transformation.
Moreover, the stable fixed point of this map lies in the open unit disc~$\mathbb{D}$,
while the unstable fixed point lies in the complementary domain $\hat{\mathbb{C}}\backslash\overline{\mathbb{D}}$.
For Eq.~(\ref{Eq:Riccati}) this means that it has exactly one stable $2\pi$-periodic solution
and this solution satisfies $|z(t)| < 1$ for all $0\le t \le 2\pi$.
\label{Proposition:Sln}
\end{proposition}

{\bf Proof:} Our proof is based on the properties of M{\"o}bius iterated function systems, described in~\cite{Vin2013}.
First, using Eq.~(\ref{Eq:Abs}) and assumption~(\ref{Ineq:c}), we obtain
$$
\df{|z|^2}{t} = 2 \Real\left( \overline{c_0(t)} z + c_1(t) + c_2(t) z \right) < 0\quad\mbox{for all}\quad z\in\mathbb{S}\quad\mbox{and}\quad t\ge 0.
$$
This means that every solution $z(t)$ of Eq.~(\ref{Eq:Riccati}) with $z(0)\in\mathbb{S}$
is trapped in the unit disc~$\mathbb{D}$, so that $|z(t)| < 1$ for all $t > 0$.
Hence, the M{\"o}bius transformation $\mathcal{M}(z)$
representing the Poincar{\'e} map of such Eq.~(\ref{Eq:Riccati}) satisfies
$|\mathcal{M}(z)| < 1$ for all $z\in\mathbb{S}$, and therefore $\mathcal{M}(\overline{\mathbb{D}})\subset\mathbb{D}$.
Due to~\cite[Theorem~4.5]{Vin2013}, the latter is a sufficient condition for $\mathcal{M}(z)$ to be a contraction on~$\mathbb{D}$.
Then, Theorem~1.1 and Theorem~3.6 from~\cite{Vin2013} imply
that $\mathcal{M}(z)$ is a hyperbolic or loxodromic M{\"o}bius transformation\footnote{Note %
that in contrast to this paper, in~\cite{Vin2013} the term loxodromic M{\"o}bius transformation
is used to denote the transformations, which are neither parabolic nor elliptic.
Therefore, hyperbolic M{\"o}bius transformations are considered there as a special case of loxodromic ones.},
that its stable fixed point lies in~$\mathbb{D}$,
and that its unstable fixed point lies outside~$\overline{\mathbb{D}}$.
Obviously, if we take the stable fixed point of~$\mathcal{M}(z)$
as the initial condition $z(0)$ for Eq.~(\ref{Eq:Riccati}),
we obtain a $2\pi$-periodic solution~$z(t)$
such that $|z(t)| < 1$ for all $0\le t \le 2\pi$.~\qed

\begin{remark}
Note that the strict inequality in~(\ref{Ineq:c}) is essential for the statement of Proposition~\ref{Proposition:Sln}.
For example, if instead of~(\ref{Ineq:c}) we use the non-strict inequality from Proposition~\ref{Proposition:D},
we cannot guarantee that the Poincar{\'e} map of Eq.~(\ref{Eq:Riccati})
is represented by a hyperbolic or loxodromic M{\"o}bius transformation only.
The corresponding counter-examples can be found in~\cite{Ome2022}.
\end{remark}

\begin{remark}
A statement similar to Proposition~\ref{Proposition:Sln} can also be formulated in the case
$$
\Real( \overline{c_0(t)} z + c_1(t) + c_2(t) z ) > 0\quad\mbox{for all}\quad z\in\mathbb{S}\quad\mbox{and}\quad 0\le t \le 2\pi.
$$
Then, the Poincar{\'e} map of Eq.~(\ref{Eq:Riccati}) is again described by a hyperbolic or loxodromic M{\"o}bius transformation.
But, the stable fixed point of this map lies outside~$\overline{\mathbb{D}}$,
while the unstable fixed point lies in~$\mathbb{D}$.
To see this, it is enough to use the substitution $z(t) = \tilde{z}(-t)$ in Eq.~(\ref{Eq:Riccati})
and consider the resulting equation as an equation with respect to $\tilde{z}(t)$.
\label{Remark:Minus}
\end{remark}

The requirements of Proposition~\ref{Proposition:Sln} are fulfilled for Eq.~(\ref{Eq:OA:Oscillators}), if $\gamma > 0$.
Indeed, inserting~(\ref{c012:Oscillators}) into~(\ref{Ineq:c}) we obtain
$$
\Real( \overline{c_0(t)} z + c_1(t) + c_2(t) z ) = - \gamma < 0.
$$
However, Proposition~\ref{Proposition:Sln} cannot be used for Eq.~(\ref{Eq:OA:Neurons}), even if $\gamma > 0$.
This follows from the fact that for the coefficients~$c_0(t)$, $c_1(t)$ and~$c_2(t)$
defined by~(\ref{c012:Neurons}) we have
$$
\Real( \overline{c_0(t)} z + c_1(t) + c_2(t) z ) = - \gamma \Real( z + 1 ).
$$
Therefore, the strict inequality~(\ref{Ineq:c}) holds for all $z\in\mathbb{S}$, except $z = -1$.
To overcome this difficulty, we propose a modified version of Proposition~\ref{Proposition:Sln}.

\begin{proposition}
Suppose
$$
\Real( \overline{c_0(t)} z + c_1(t) + c_2(t) z ) < 0\quad\mbox{for all}\quad z\in\mathbb{S}\backslash z_*\quad\mbox{and}\quad 0\le t \le 2\pi,
$$
where $z_*$ is some point on the unit circle~$\mathbb{S}$.
Moreover, suppose that the solution of Eq.~(\ref{Eq:Riccati}) with the initial condition $z(0) = z_*$
exists for $-2\pi \le t \le 0$ and satisfies $|z(-2\pi)|\ne 1$.
Then, all the conclusions of Proposition~\ref{Proposition:Sln} remain valid.
\label{Proposition:Sln_}
\end{proposition}

{\bf Proof:} We only need to show that $\mathcal{M}(\mathbb{S})\subset\mathbb{D}$,
then we can repeat the remaining arguments of the proof of Proposition~\ref{Proposition:Sln}.
For this, we consider a general solution $z(t)$ of Eq.~(\ref{Eq:Riccati}) with $z(0)\in\mathbb{S}$.
In this case, Proposition~\ref{Proposition:D} ensures that $z(t)\in\overline{\mathbb{D}}$ for all $t > 0$.
Could it be that $z(2\pi)\in\mathbb{S}$? The answer is no, because
$$
\df{|z|^2}{t} = 2 \Real\left( \overline{c_0(t)} z + c_1(t) + c_2(t) z \right) < 0\quad\mbox{for all}\quad z\in\mathbb{S}\backslash z_*\quad\mbox{and}\quad t = 2\pi,
$$
and because of the assumption
that the solution of Eq.~(\ref{Eq:Riccati}) with $z(2\pi) = z_*$ satisfies $z(0)\notin\mathbb{S}$.
This ends the proof.~\qed

\section{Solution of periodic complex Riccati equation}
\label{Sec:Moebius}

The complex Riccati equation~(\ref{Eq:Riccati}) is a nonlinear differential equation that cannot be solved analytically.
Therefore, its periodic solutions can usually be found only by numerical methods,
such as the shooting method or the collocation method.
However, without a good initial guess, each of these methods may involve
a large number of iterations and thus require a long computational time
to ensure the desired accuracy of the result.
A more efficient way to calculate the periodic solutions of Eq.~(\ref{Eq:Riccati})
can be proposed using the relation between the M{\"o}bius transformation
and the Poincar{\'e} map of Eq.~(\ref{Eq:Riccati}).

Suppose that the assumption of Proposition~\ref{Proposition:Sln} is satisfied.
Then, by choosing three distinct points $z_k\in\overline{\mathbb{D}}$, $k=1,2,3$,
and solving Eq.~(\ref{Eq:Riccati}) with different initial conditions $z(0) = z_k$,
we obtain three complex functions $U_k(t)$.
Due to Proposition~\ref{Proposition:D} each of these functions is bounded and satisfies $|U_k(t)|\le 1$.
Let us denote $w_k = U_k(2\pi)$, then
\begin{equation}
w_k = \mathcal{M}(z_k),\quad k=1,2,3,
\label{z_to_w}
\end{equation}
where $\mathcal{M}(z)$ is a M{\"o}bius transformation of the form~(\ref{Def:M})
that represents the Poincar{\'e} map of Eq.~(\ref{Eq:Riccati}).
It is well-known~\cite{Book:Needham} that the information contained in the relations~(\ref{z_to_w})
is sufficient to determine the coefficients~$a$, $b$, $c$ and $d$ in the expression of~$\mathcal{M}(z)$.
The corresponding explicit formulas read
\begin{eqnarray*}
&&
a = \det \left(
\begin{array}{ccc}
 z_1 w_1 & w_1 & 1 \\[2mm]
 z_2 w_2 & w_2 & 1 \\[2mm]
 z_3 w_3 & w_3 & 1
\end{array}
\right),
\qquad
b = \det \left(
\begin{array}{ccc}
 z_1 w_1 & z_1 & w_1 \\[2mm]
 z_2 w_2 & z_2 & w_2 \\[2mm]
 z_3 w_3 & z_3 & w_3
\end{array}
\right),
\\[2mm]
&&
c = \det \left(
\begin{array}{ccc}
 z_1 & w_1 & 1 \\[2mm]
 z_2 & w_2 & 1 \\[2mm]
 z_3 & w_3 & 1
\end{array}
\right),
\qquad\phantom{w_1}
d = \det \left(
\begin{array}{ccc}
 z_1 w_1 & z_1 & 1 \\[2mm]
 z_2 w_2 & z_2 & 1 \\[2mm]
 z_3 w_3 & z_3 & 1
\end{array}
\right).
\end{eqnarray*}
Once the transformation~$\mathcal{M}(z)$ is defined,
its fixed points can be found by solving the equation
$$
z = \fr{a z + b}{c z + d}.
$$
The latter is obviously equivalent to the quadratic equation
$$
c z^2 + d z - a z - b = 0
$$
and has two roots
$$
z_\pm = \fr{a - d \pm \sqrt{ (a - d)^2 + 4 b c }}{2 c}.
$$
Now both periodic solutions of Eq.~(\ref{Eq:Riccati}) can be obtained
by integrating this equation with the initial conditions $z(0) = z_-$ and $z(0) = z_+$.
Importantly, Proposition~\ref{Proposition:Sln} ensures
that one of the fixed points~$z_-$ or $z_+$ lies in the open unit disc~$\mathbb{D}$
and the corresponding periodic solution satisfies $|z(t)| < 1$.

\begin{remark}
In some cases, the map trajectories of the M{\"o}bius transformation~$\mathcal{M}(z)$
converge to its fixed point in~$\mathbb{D}$ so fast
that the numbers~$w_1$, $w_2$ and~$w_3$ lie extremely close to each other.
Then, the above formulas for the coefficients~$a$, $b$, $c$ and~$d$ cannot be applied,
due to the vanishing values of the determinants.
In this case, the fixed point of~$\mathcal{M}(z)$ that lies in~$\mathbb{D}$
can simply be approximated by the mean $(w_1 + w_2 + w_3)/3$.
\end{remark}

\section{Application to travelling chimera states}
\label{Sec:Chimera}

In~\cite{Ome2020}, the author of this paper considered moving coherence-incoherence patterns,
called travelling chimera states, in a ring network of nonlocally coupled phase oscillators.
In the continuum limit of infinitely many oscillators,
the long-term dynamics of such a network is described by an integro-differential equation~\cite{Lai2009,Ome2018}
\begin{equation}
\df{z}{t} = - \gamma z + \fr{1}{2} e^{-i \alpha} \mathcal{G} z - \fr{1}{2} e^{i \alpha} z^2 \mathcal{G} \overline{z},
\label{Eq:OA}
\end{equation}
where $z(x,t)$ is the unknown complex-valued function
satisfying the periodic boundary condition $z(x+2\pi,t) = z(x,t)$,
$\gamma > 0$ is a parameter analogous to the width of the Lorentzian distribution in Section~\ref{Sec:Models}, and
$$
(\mathcal{G} \varphi)(x) = \int_{-\pi}^\pi G( x - y ) \varphi(y) dy
$$
is a convolution integral operator with a non-constant real $2\pi$-periodic kernel $G(x)$.
Moreover, each travelling chimera state corresponds in Eq.~(\ref{Eq:OA}) to a travelling wave solution of the form
\begin{equation}
z(x,t) = a(x - s t) e^{i \Omega t}
\label{Eq:Ansatz}
\end{equation}
where $a(x)\in\mathbb{D}$ is the wave profile, $s\in\mathbb{R}$ is the wave speed
and $\Omega\in\mathbb{R}$ is its complex-phase velocity.
The analysis of travelling waves~(\ref{Eq:Ansatz}) carried out in~\cite{Ome2020}
revealed unexpected oscillatory properties of their profiles~$a(x)$
and a non-monotonic dependence of the speed $s$ on the system parameters.
Below, we extend this analysis, using the advantages
provided by the semi-analytic description of the periodic solutions
of the complex Riccati equation~(\ref{Eq:Riccati}).

\subsection{Self-consistency equation}

In this section, we derive a self-consistency equation for travelling waves~(\ref{Eq:Ansatz}),
which helps to significantly speed up their numerical calculation.
Inserting ansatz~(\ref{Eq:Ansatz}) into Eq.~(\ref{Eq:OA}) we obtain
\begin{equation}
i \Omega a - s a' = - \gamma a + \fr{1}{2} e^{-i \alpha} \mathcal{G} a - \fr{1}{2} e^{i \alpha} a^2 \mathcal{G} \overline{a},
\label{Eq:a:old}
\end{equation}
where $a(x)$ is the unknown complex-valued function and $a'$ is its usual derivative.
Next, subtracting $i \Omega a$ in the both sides and dividing the resulting equation by $(- s)\ne 0$, we get
$$
a' = \fr{\gamma + i \Omega}{s} a + \fr{1}{(-2 s)} e^{-i \alpha} \mathcal{G} a - \fr{1}{(-2 s)} e^{i \alpha} a^2 \mathcal{G} \overline{a}.
$$
The latter equation can be written in the form
\begin{equation}
a' = w(x) + \zeta a - \overline{w(x)} a^2
\label{Eq:a}
\end{equation}
where
\begin{equation}
\zeta = \fr{\gamma + i \Omega}{s}\qquad\mbox{and}\qquad
w(x) = - \fr{1}{2 s} e^{-i \alpha} \mathcal{G} a.
\label{Def:zeta:w}
\end{equation}
This is the complex Riccati equation~(\ref{Eq:Riccati})
with $c_0(x) = w(x)$, $c_1(x) = \zeta$ and $c_2(x) = - \overline{w(x)}$.
Obviously, for $\gamma > 0$ and $s\ne 0$ we have
$$
\Real( \overline{c_0(x)} z + c_1(x) + c_2(x) z ) =
\Real( \overline{w(x)} z + \zeta - \overline{w(x)} z ) = \Real\:\zeta = \gamma / s \ne 0,
$$
therefore for every $w\in C_\mathrm{per}([0,2\pi];\mathbb{C})$ and $\zeta\in\{z\in\mathbb{C}\::\:\Real\:z \ne 0\}$
there exists a unique $2\pi$-periodic solution of Eq.~(\ref{Eq:a}) that lies entirely in the unit disc~$\mathbb{D}$,
see Proposition~\ref{Proposition:Sln} and Remark~\ref{Remark:Minus}.
(Note that although for $s > 0$ the above solution $a(x)$ is unstable with respect to Eq.~(\ref{Eq:a}),
the corresponding travelling wave~(\ref{Eq:Ansatz}) may be stable or unstable with respect to Eq.~(\ref{Eq:OA}).)
In the following, we denote the obtained solution operator for Eq.~(\ref{Eq:a}) by $\mathcal{U}(w,\zeta)$.
More precisely, this operator determines a mapping
$$
\mathcal{U}\::\:(w,\zeta)\in C_\mathrm{per}([0,2\pi];\mathbb{C})\times\{z\in\mathbb{C}\::\:\Real\:z \ne 0\}\mapsto a\in C^1_\mathrm{per}([0,2\pi];\mathbb{D}).
$$
(Note that in the above definition we have $|a(t)| < 1$ for all $t\in[0,2\pi]$!)
If $a = \mathcal{U}(w,\zeta)$, then to agree with the formulas~(\ref{Def:zeta:w}) we must have
$$
w = - \fr{1}{2 s} e^{-i \alpha} \mathcal{G} \mathcal{U}(w,\zeta),
$$
or equivalently
\begin{equation}
-2 s e^{i \alpha} w = \mathcal{G} \mathcal{U}\left( w, \fr{\gamma + i \Omega}{s} \right).
\label{Eq:SC}
\end{equation}

\begin{remark}
Every travelling wave solution of Eq.~(\ref{Eq:OA}) is not uniquely determined.
Given a wave profile $a_0(x)$, one obtains infinitely many other solutions of Eq.~(\ref{Eq:OA})
by the formula $a(x) = a_0(x-\xi) e^{i \phi}$ with $\xi, \phi\in\mathbb{R}$.
Moreover, it is easy to verify that the same symmetry property is inherited by Eq.~(\ref{Eq:SC}).
In other words, if $w_0(x)$ solves Eq.~(\ref{Eq:SC}), then $w(x) = w_0(x-\xi) e^{i \phi}$
with $\xi, \phi\in\mathbb{R}$ is a solution of Eq.~(\ref{Eq:SC}) too.
\label{Remark:Invariance}
\end{remark}

Eq.~(\ref{Eq:SC}) is the self-consistency equation to be solved with respect to
$w\in C_\mathrm{per}([-\pi,\pi];\mathbb{C})$, $s\in\mathbb{R}\backslash\{0\}$ and $\Omega\in\mathbb{R}$.
A unique solution of Eq.~(\ref{Eq:SC}) is fixed by two supplementary pinning conditions
\begin{eqnarray}
&&
\Imag\left( \int_{-\pi}^\pi w(x) dx \right) = 0,
\label{Eq:Pinning:1}\\[2mm]
&&
\Imag\left( \int_{-\pi}^\pi w(x) e^{-i x} dx \right) = 0.
\label{Eq:Pinning:2}
\end{eqnarray}

\subsection{Self-consistency equation for a trigonometric coupling function}

Now we consider travelling wave solutions of Eq.~(\ref{Eq:OA}) in the case of a trigonometric coupling function
\begin{equation}
G(x) = \fr{1}{2\pi} \left( 1 + A \cos x + B \sin x \right)
= \fr{1}{2\pi} \left( 1 + \fr{A - B i}{2} e^{i x} + \fr{A + B i}{2} e^{-i x} \right)
\label{Def:G}
\end{equation}
with two real coefficients $A$ and $B$.
The integral operator $\mathcal{G}$ with such a kernel $G(x)$ is a finite-rank operator.
This follows from its representation formula
\begin{equation}
\mathcal{G} u = \langle u, \psi_1 \rangle + \fr{A - B i}{2} \langle u, \psi_2 \rangle e^{i x} + \fr{A + B i}{2} \langle u, \psi_3 \rangle e^{-i x}
\label{Eq:Representation}
\end{equation}
where
$$
\langle \phi, \psi \rangle = \fr{1}{2\pi} \int_{-\pi}^\pi \phi(x) \overline{\psi(x)} dx
$$
denotes the usual $L^2$-scalar product, and
$$
\psi_1(x) = 1,\quad \psi_2(x) = e^{i x},\quad \psi_3(x) = e^{-i x}
$$
are three basis functions that span the image of the operator~$\mathcal{G}$.
Formula~(\ref{Eq:Representation}) implies that every solution of Eq.~(\ref{Eq:SC})
with the above trigonometric coupling function $G(x)$ has the form
\begin{equation}
w(x) = \hat{w}_0 + \hat{w}_1 e^{i x} + \hat{w}_2 e^{-i x}
\label{Def:w:general}
\end{equation}
where $\hat{w}_0, \hat{w}_1, \hat{w}_2\in\mathbb{C}$.
Moreover, if the pinning conditions~(\ref{Eq:Pinning:1}) and~(\ref{Eq:Pinning:2}) are satisfied,
then $\hat{w}_0$ and $\hat{w}_1$ must be real.
Note that due to Remark~\ref{Remark:Invariance}, we can always achieve this
by using two continuous symmetries of the solutions of Eq.~(\ref{Eq:SC}).
Indeed, suppose that $w(x)$ is a solution of Eq.~(\ref{Eq:SC}) in the most general form~(\ref{Def:w:general}).
Using the complex-phase-shift symmetry, we can obtain another solution
$$
w(x) = |\hat{w}_0| + \fr{\overline{\hat{w}_0}}{|\hat{w}_0|}\hat{w}_1 e^{i x} + \fr{\overline{\hat{w}_0}}{|\hat{w}_0|}\hat{w}_2 e^{-i x}
$$
with real first coefficient. Then, using the translation symmetry we can get a solution of the form
$$
w(x) = |\hat{w}_0| + |\hat{w}_1| e^{i x} + \fr{\overline{\hat{w}_0}^2}{|\hat{w}_0|^2}\fr{\hat{w}_1}{|\hat{w}_1|} \hat{w}_2 e^{-i x},
$$
which has real first and second coefficients and thus satisfies~(\ref{Eq:Pinning:1}) and~(\ref{Eq:Pinning:2}).

In the rest of this section we show how the system~(\ref{Eq:SC})--(\ref{Eq:Pinning:2}),
which determines travelling wave solutions of Eq.~(\ref{Eq:OA}),
can be reduced to a finite-dimensional nonlinear system.
For this, we look for solutions of Eq.~(\ref{Eq:SC}) in the following form
$$
w(x) = p + q e^{i x} + \hat{w}_* e^{-i x}
$$
where $p,q\in\mathbb{R}$ and $\hat{w}_*\in\mathbb{C}$.
Such ansatz ensures that the pinning conditions~(\ref{Eq:Pinning:1}) and~(\ref{Eq:Pinning:2}) are satisfied automatically.
Then, from formula~(\ref{Eq:Representation}) and from the linear independence of functions~$\psi_k(x)$
it follows that Eq.~(\ref{Eq:SC}) is equivalent to a system of three scalar complex equations:
\begin{eqnarray}
-2 s e^{i \alpha} p &=& \left\langle \mathcal{U}\left( w, \fr{\gamma + i \Omega}{s} \right), \psi_1 \right\rangle,\label{System:1}\\[2mm]
-2 s e^{i \alpha} q &=& \fr{A - B i}{2} \left\langle \mathcal{U}\left( w, \fr{\gamma + i \Omega}{s} \right), \psi_2 \right\rangle,\label{System:2}\\[2mm]
-2 s e^{i \alpha} \hat{w}_* &=& \fr{A + B i}{2} \left\langle \mathcal{U}\left( w, \fr{\gamma + i \Omega}{s} \right), \psi_3 \right\rangle.\label{System:3}
\end{eqnarray}
This system can be solved with respect to six unknowns $p$, $q$, $\Real\:\hat{w}_*$, $\Imag\:\hat{w}_*$, $s$ and $\Omega$,
using a standard Newton's method. Given a system's solution, the corresponding wave profile can be calculated by the formula
$$
a(x) = \mathcal{U}\left( w(x), \fr{\gamma + i \Omega}{s} \right).
$$

\subsection{Results}

Using system~(\ref{System:1})--(\ref{System:3}),
we carried out a detailed analysis of travelling wave solutions of Eq.~(\ref{Eq:OA})
in the case of trigonometric coupling~(\ref{Def:G}) with $B\ne 0$.
We kept the parameters $A = 0.9$ and $\alpha = \pi/2 - 0.1$ fixed
and used a pseudo-arclength continuation to follow
three branches of travelling waves for $\gamma = 0.005$, $0.01$ and $0.02$.
For each travelling wave we calculated its twist,
which is defined as the net number of multiples of $2\pi$
through which the argument of the complex wave profile $a(x)$ decreases
as the spatial domain is traversed once.
Moreover, the stability of travelling waves was determined
by the linearization of Eq.~(\ref{Eq:OA}) about the found solution,
which led to the consideration of the eigenvalue problem~\cite{Ome2020}
\begin{eqnarray}
\lambda v_+ &=& s \partial_x v_+ - \eta(x) v_+ + \fr{1}{2} e^{-i \alpha} \mathcal{G} v_+ - \fr{1}{2} e^{i \alpha} a^2(x) \mathcal{G} v_-, \\[2mm]
\lambda v_- &=& s \partial_x v_- - \overline{\eta}(x) v_- + \fr{1}{2} e^{i \alpha} \mathcal{G} v_- - \fr{1}{2} e^{-i \alpha} \overline{a}^2(x) \mathcal{G} v_+,
\end{eqnarray}
where $a(x)$, $s$ and $\Omega$ are the wave profile, speed and complex-phase velocity of the reference travelling wave,
$\lambda$ and $(v_+(x),v_-(x))^\mathrm{T}$ are the eigenvalue and eigenfunction, and
$$
\eta(x) = \gamma + i \Omega + e^{i \alpha} a(x) \mathcal{G} \overline{a}.
$$
The above integral system was discretized on a uniform grid of $1000$ points
and solved as a matrix eigenvalue problem.

Note that in contrast to the method based on the Lyapunov-Schmidt reduction of Eq.~(\ref{Eq:a:old})
and used in~\cite{Ome2020} to calculate the travelling wave solutions of Eq.~(\ref{Eq:OA}),
the numerical method proposed in this paper allowed us to perform the same calculations much faster.
Therefore, we could calculate complete solution branches,
also in the case of values $\gamma$ smaller and larger
than the value considered in~\cite{Ome2020}.
Next we describe the obtained results.
%%%%%%%%%%%%%%%%%%%%%%%%%%%%%%%%%%%%%%%
\begin{figure}[hb!]
\begin{minipage}[c]{0.26\textwidth}
\includegraphics[height=\textwidth,angle=270]{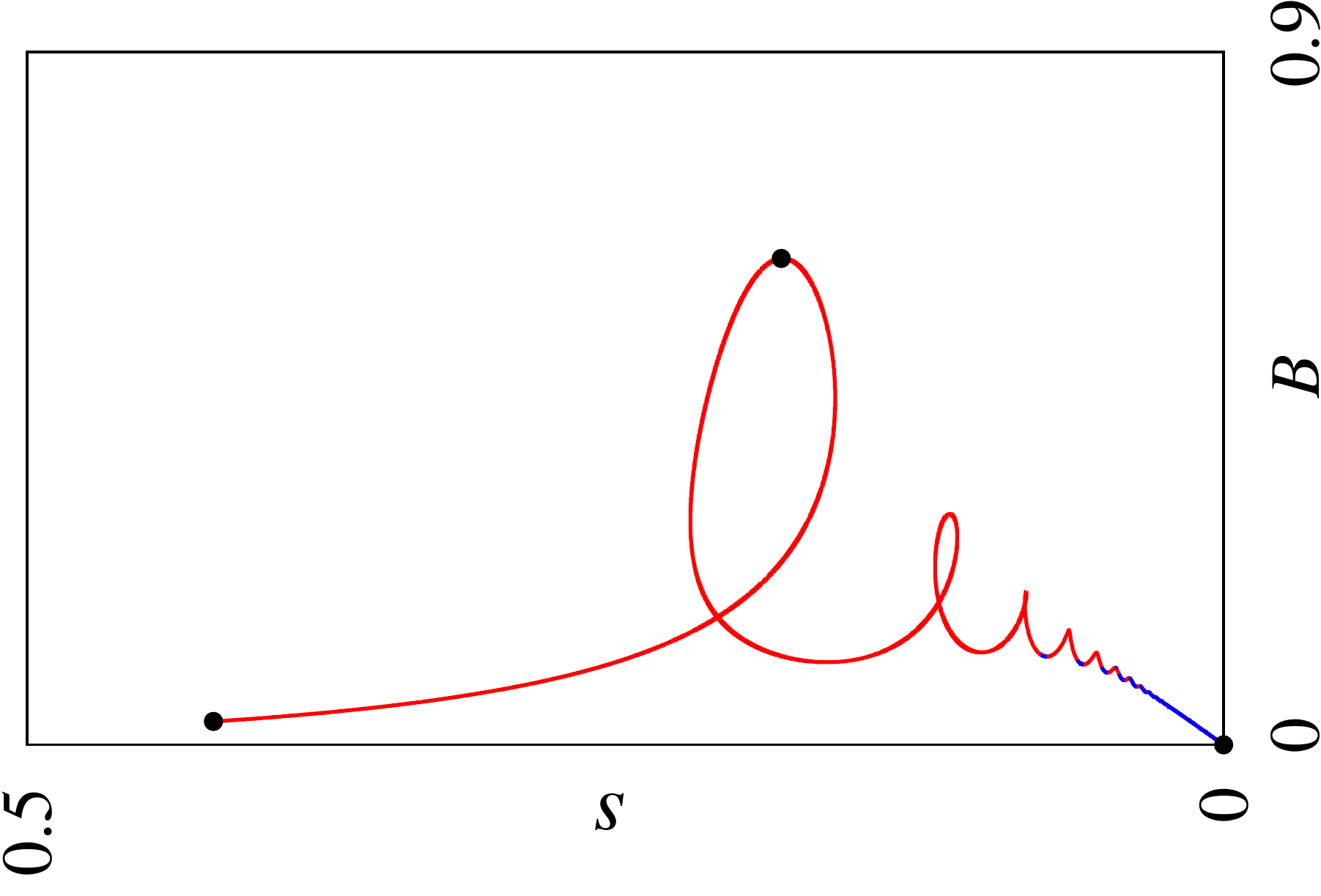}
\end{minipage}
\hspace{0.02\textwidth}
\begin{minipage}[c]{0.26\textwidth}
\includegraphics[height=\textwidth,angle=270]{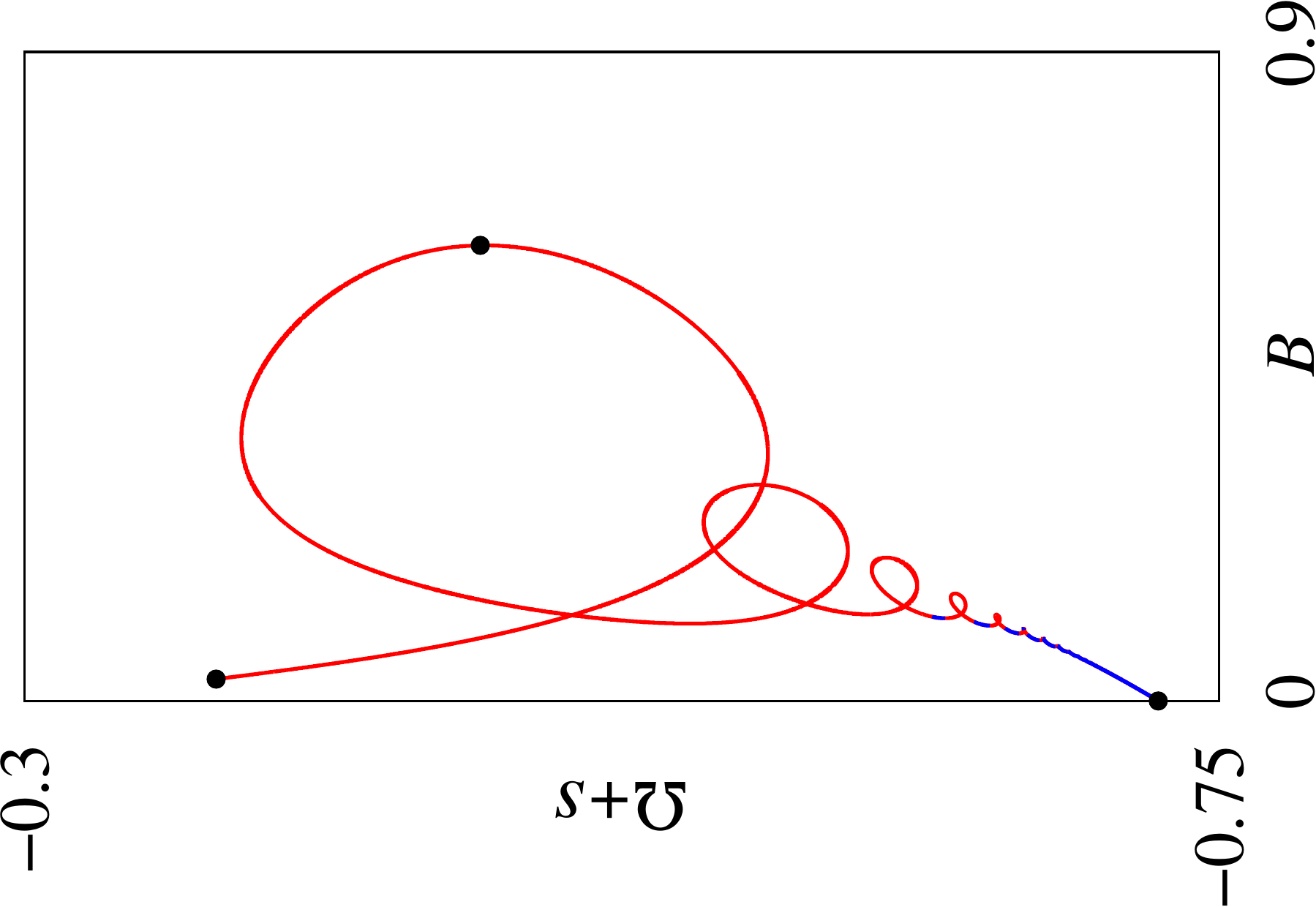}
\end{minipage}
\hspace{0.02\textwidth}
\begin{minipage}[c]{0.4\textwidth}
\includegraphics[height=\textwidth,angle=270]{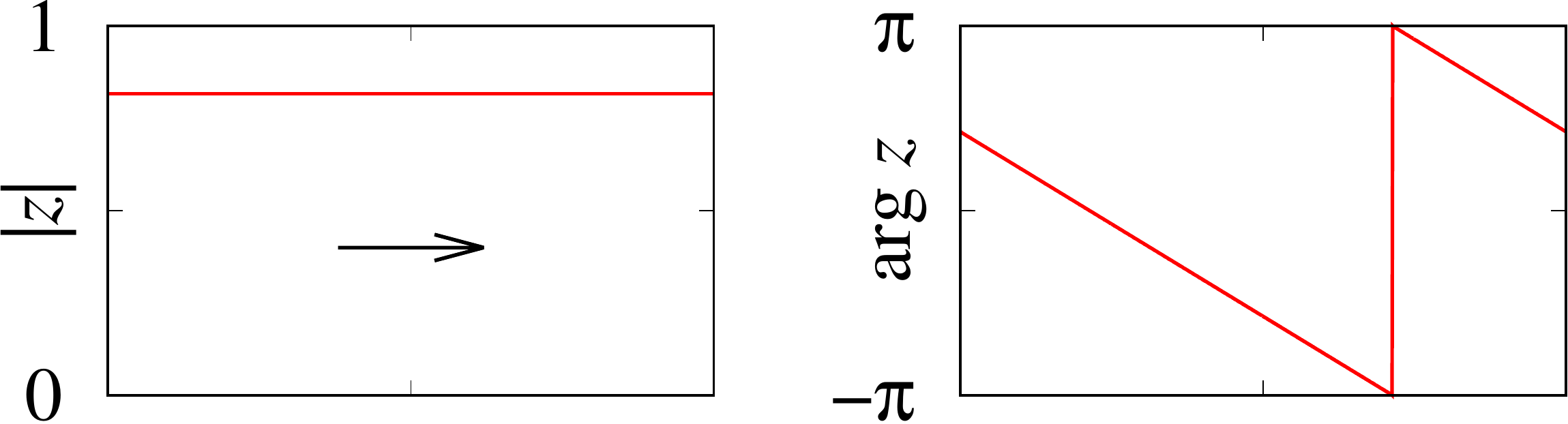}\\[2mm]
\includegraphics[height=\textwidth,angle=270]{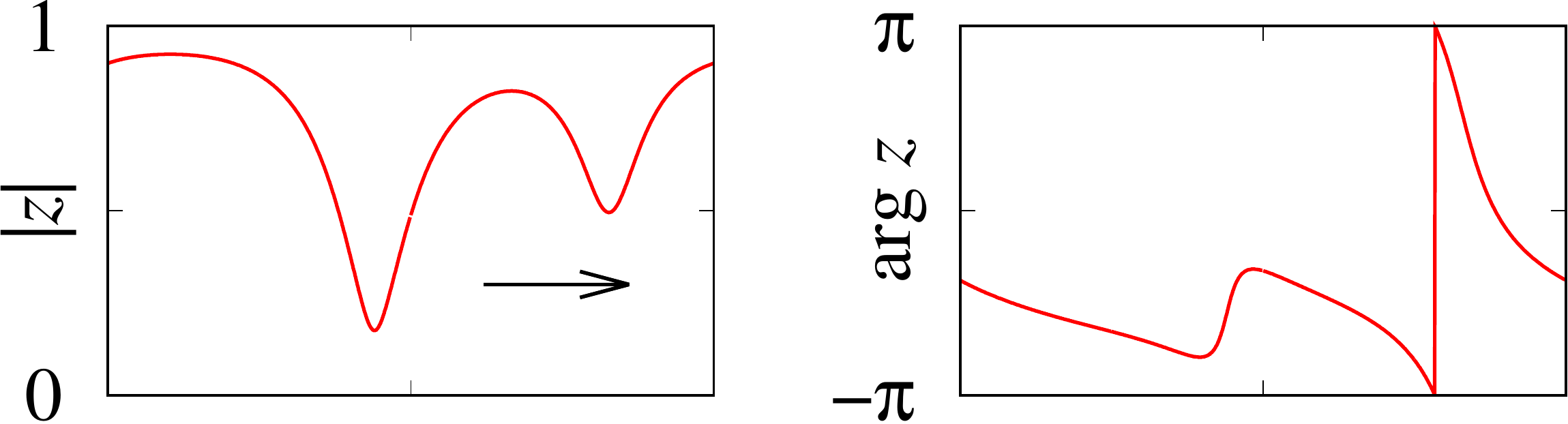}\\[2mm]
\includegraphics[height=1.01\textwidth,angle=270]{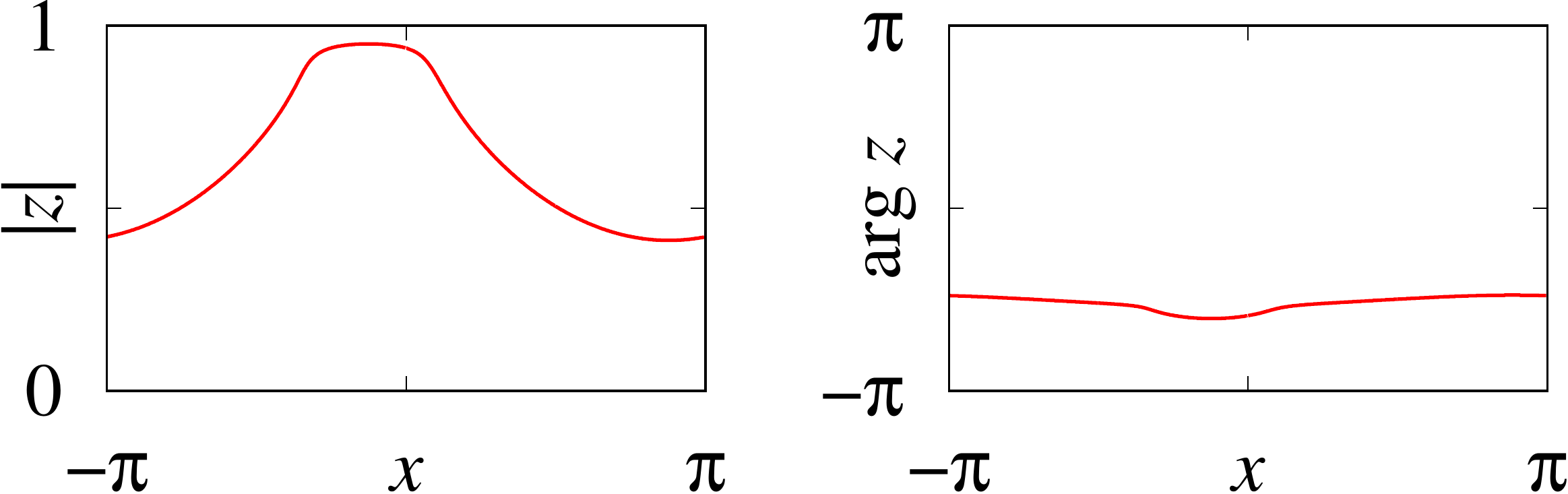}
\end{minipage}
\caption{Speed ($s$) and complex-phase velocity ($\Omega$) of the travelling wave for $\gamma = 0.01$.
Blue: stable, red: unstable. The three right panels show the wave profiles $a(x)$,
which correspond to the points indicated by dots on the left and middle panels.
Arrows on the $|z|$-graphs show the direction of the wave motion.}
\label{Fig:Diagram:0_01}
\end{figure}
%%%%%%%%%%%%%%%%%%%%%%%%%%%%%%%%%%%%%%%

\begin{figure}[ht!]
\begin{minipage}[c]{0.26\textwidth}
\includegraphics[height=\textwidth,angle=270]{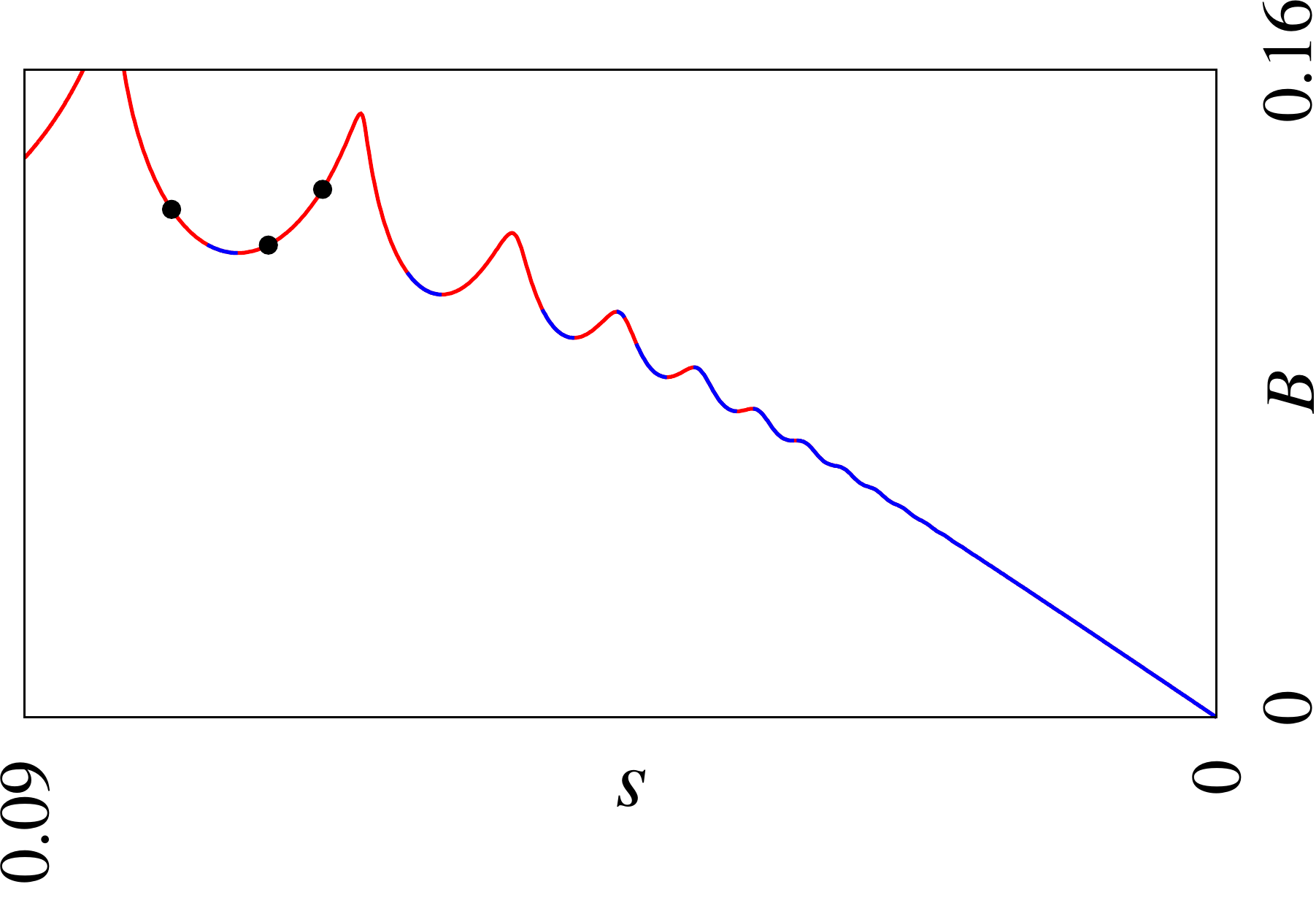}
\end{minipage}
\hspace{0.02\textwidth}
\begin{minipage}[c]{0.26\textwidth}
\includegraphics[height=\textwidth,angle=270]{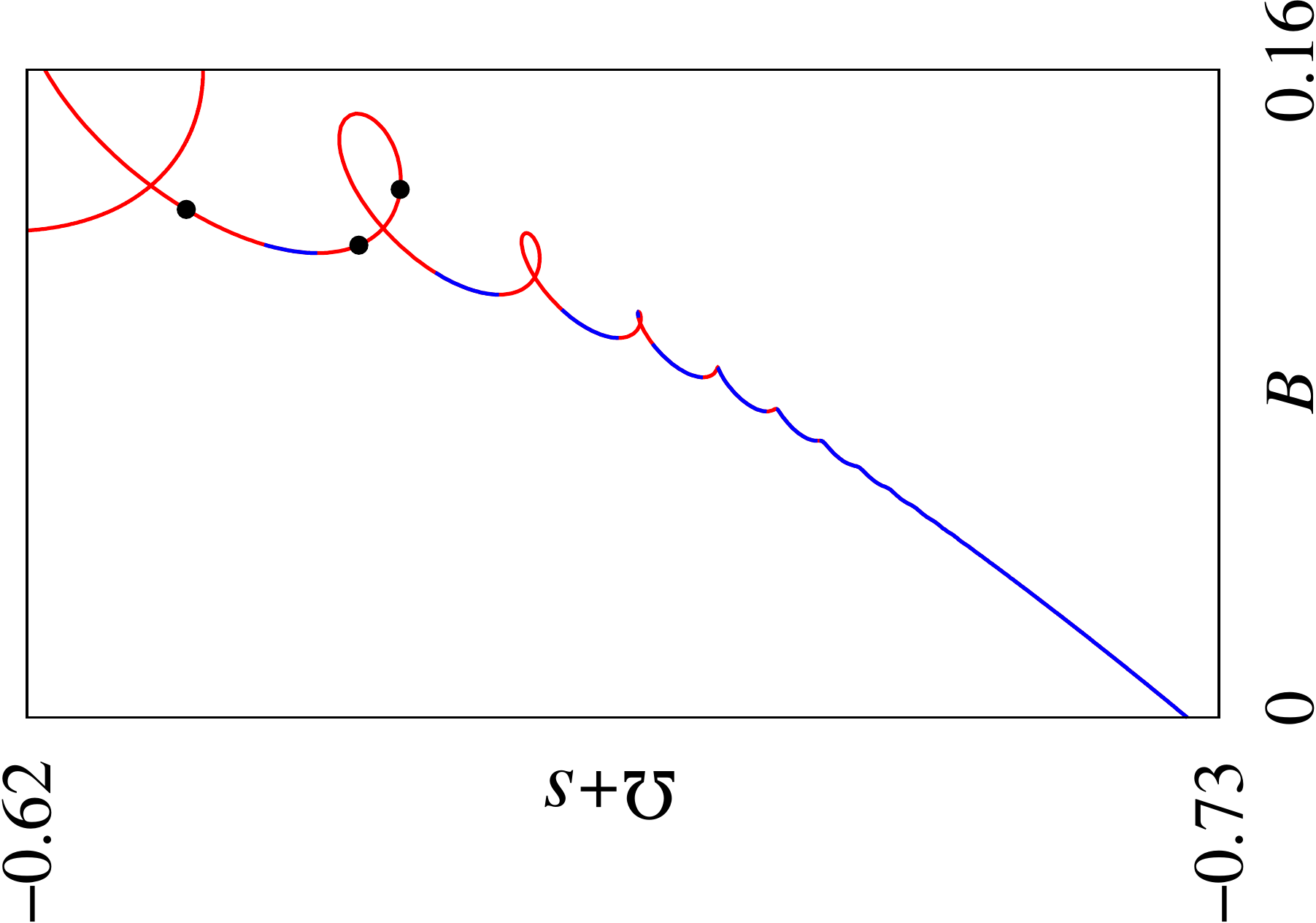}
\end{minipage}
\hspace{0.02\textwidth}
\begin{minipage}[c]{0.4\textwidth}
\includegraphics[height=\textwidth,angle=270]{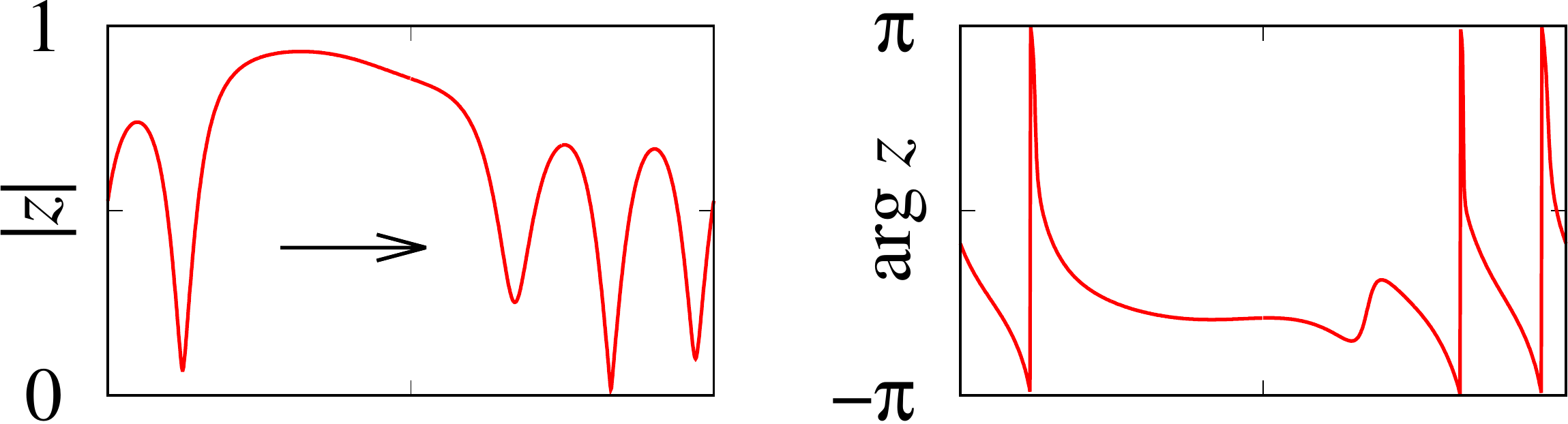}\\[2mm]
\includegraphics[height=\textwidth,angle=270]{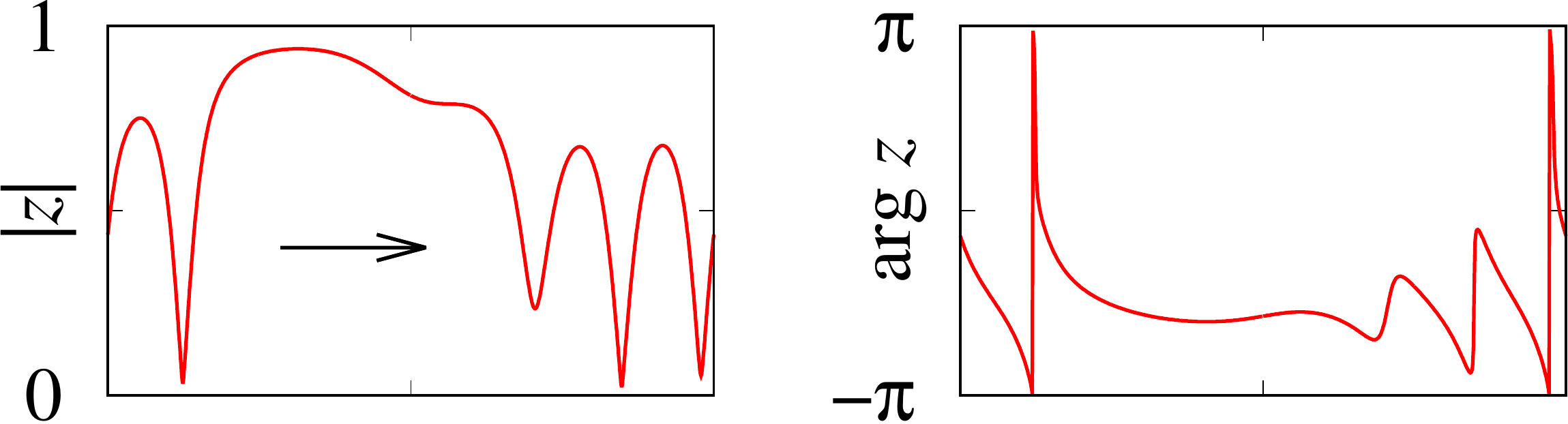}\\[2mm]
\includegraphics[height=1.01\textwidth,angle=270]{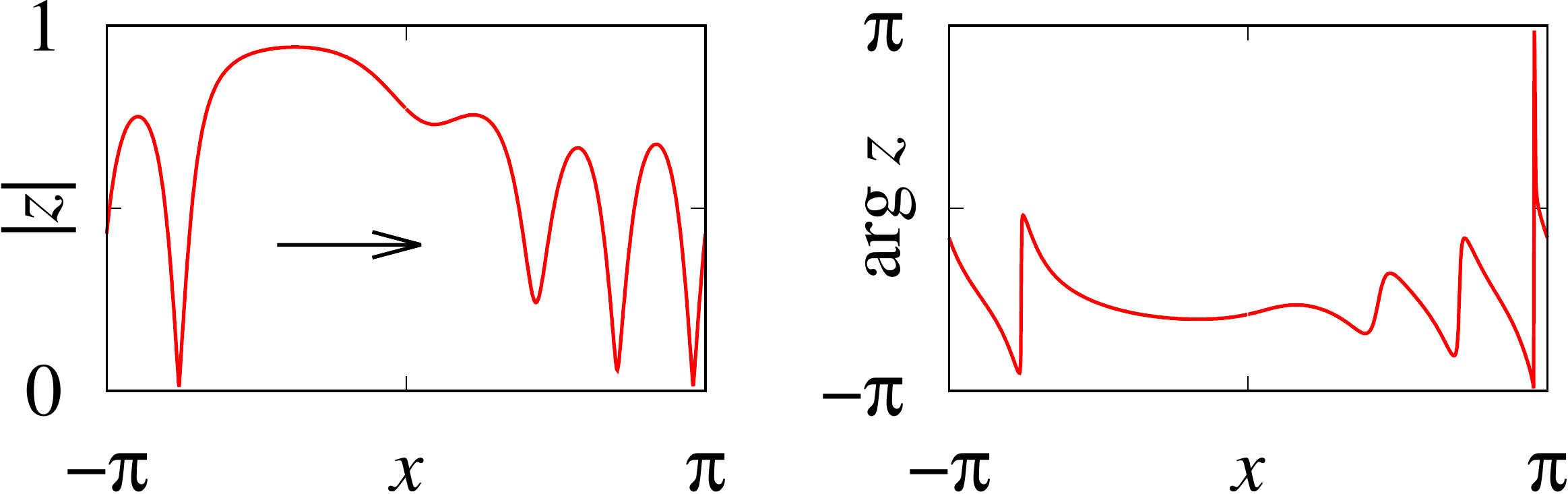}
\end{minipage}
\caption{A zoom of Fig.~\ref{Fig:Diagram:0_01}.
The data are identical to those shown in Fig.~4 in~\cite{Ome2020}.
But here they were computed with the new continuation method based on Eq.~(\ref{Eq:SC}).
Other notations the same as in Fig.~\ref{Fig:Diagram:0_01}.}
\label{Fig:Diagram:0_01_}
\end{figure}

For $\gamma = 0.01$, the branch of travelling waves starts out as twist-$0$ for $B\approx 0$,
see Fig.~\ref{Fig:Diagram:0_01}.
It remains twist-$0$ for small speeds $s$, but then it becomes twist-$1$ for $s\in(0.0665,0.0683)$,
twist-$2$ for $s\in(0.0683,0.0749)$, and twist-$3$ for $s\in(0.0749,0.0827)$, as explained in Fig.~\ref{Fig:Diagram:0_01_}.
For larger values of $s$, the inverse order of transformations is observed:
from twist-$3$ into twist-$2$ and then into twist-$1$.
It is noteworthy that stable travelling waves occur with relatively low speeds only,
while all waves with $s > 0.0762$ are unstable.
Moreover, the endpoint of the branch, i.e. the wave with the largest speed $s_\mathrm{max}\approx 0.42$, corresponds to a splay state
\begin{equation}
z(x,t) = a_\mathrm{max} e^{-i (x - s_\mathrm{max} t)} e^{i \Omega_\mathrm{max} t} = a_\mathrm{max} e^{-i x} e^{i ( \Omega_\mathrm{max} + s_\mathrm{max} ) t}
\label{Eq:Splay}
\end{equation}
with $a_\mathrm{max}\approx 0.82$ and $\Omega_\mathrm{max}\approx -0.37$.
Remark that the exponent in the right-hand side of formula~(\ref{Eq:Splay}) is our motivation to show the sum $\Omega + s$
instead of the complex-phase velocity~$\Omega$ in Figs.~\ref{Fig:Diagram:0_01}, \ref{Fig:Diagram:0_01_} and later.
%%%%%%%%%%%%%%%%%%%%%%%%%%%%%%%%%%%%%%%
%%%%%%%%%%%%%%%%%%%%%%%%%%%%%%%%%%%%%%%

\begin{figure}[ht!]
\begin{minipage}[c]{0.26\textwidth}
\includegraphics[height=\textwidth,angle=270]{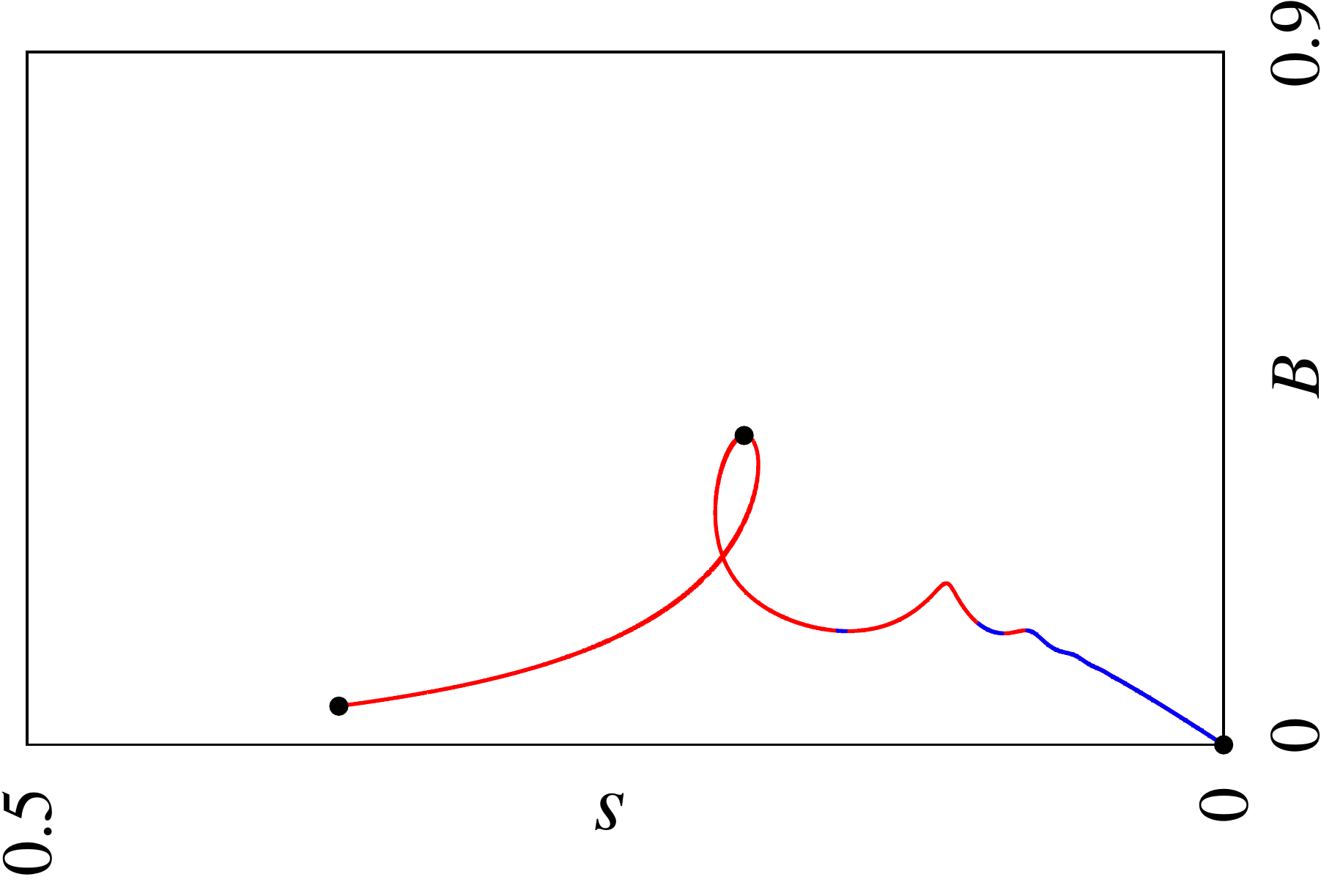}
\end{minipage}
\hspace{0.02\textwidth}
\begin{minipage}[c]{0.26\textwidth}
\includegraphics[height=\textwidth,angle=270]{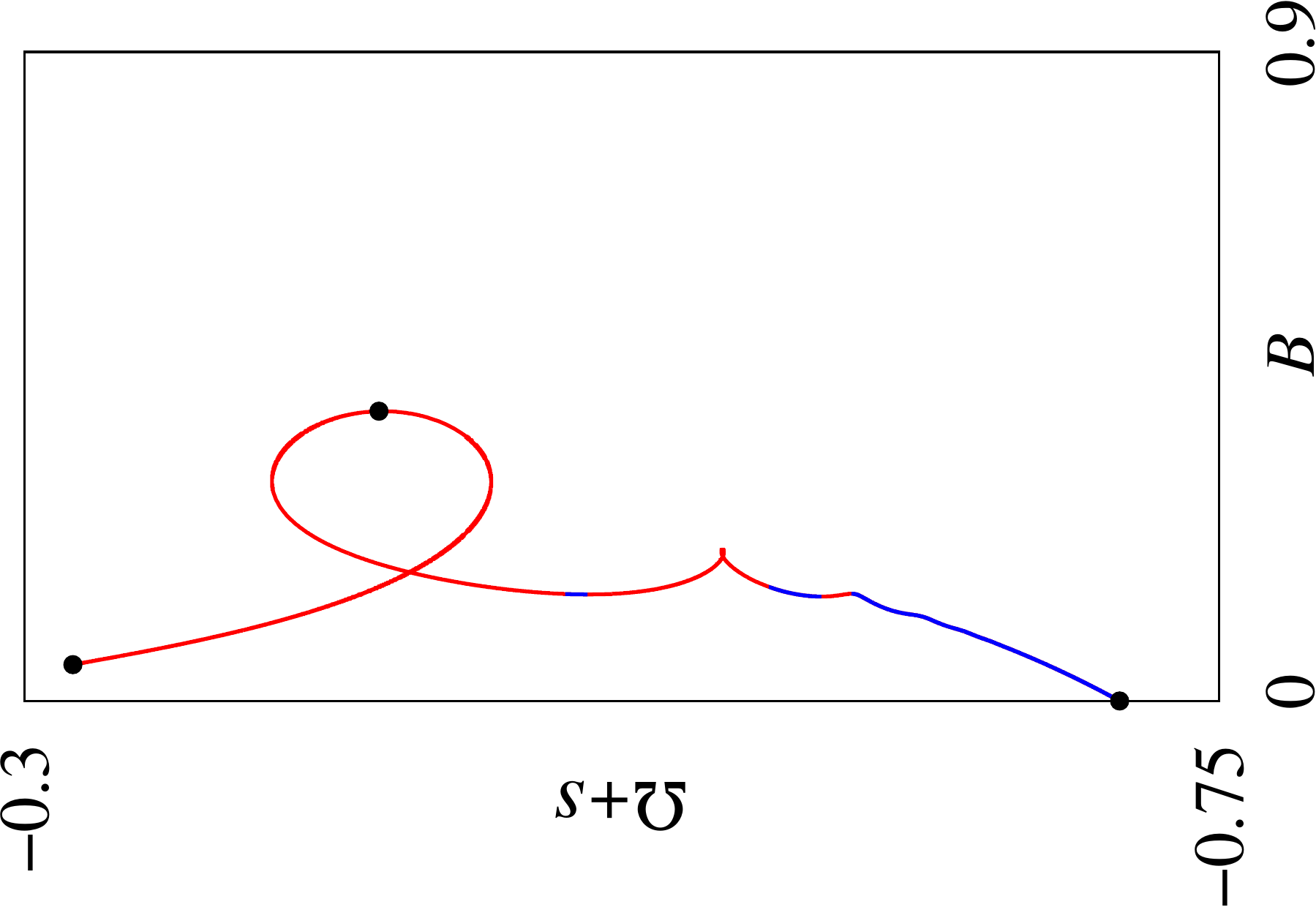}
\end{minipage}
\hspace{0.02\textwidth}
\begin{minipage}[c]{0.4\textwidth}
\includegraphics[height=\textwidth,angle=270]{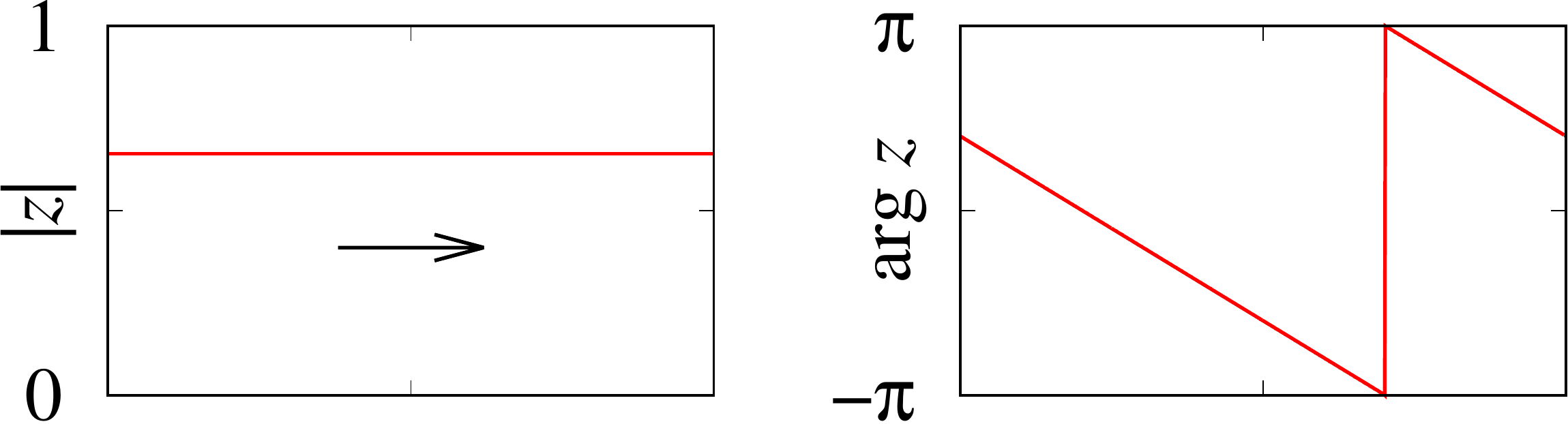}\\[2mm]
\includegraphics[height=\textwidth,angle=270]{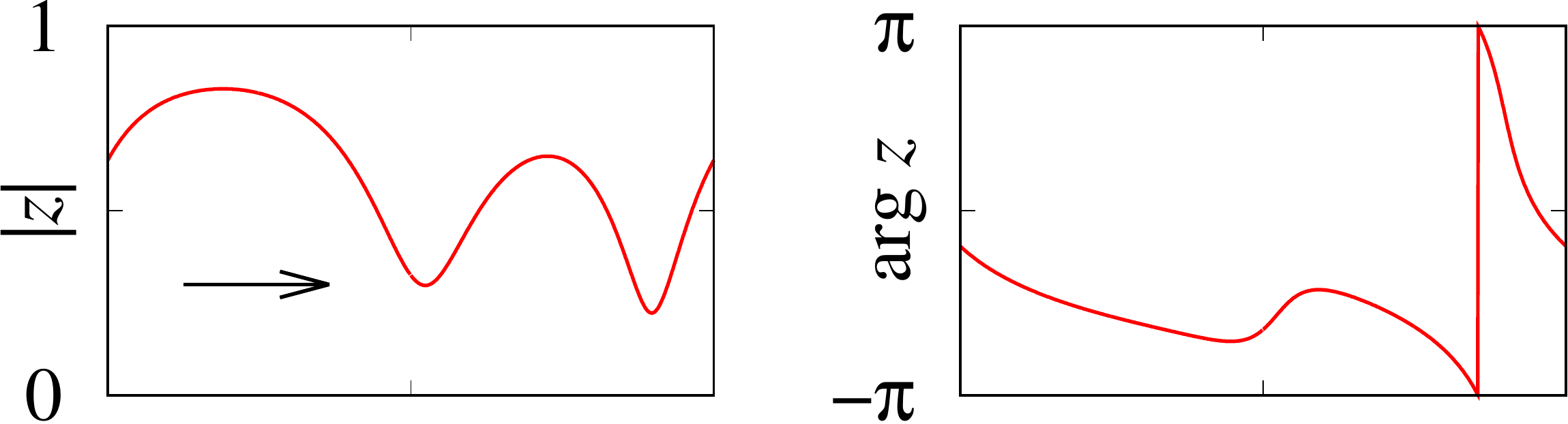}\\[2mm]
\includegraphics[height=1.01\textwidth,angle=270]{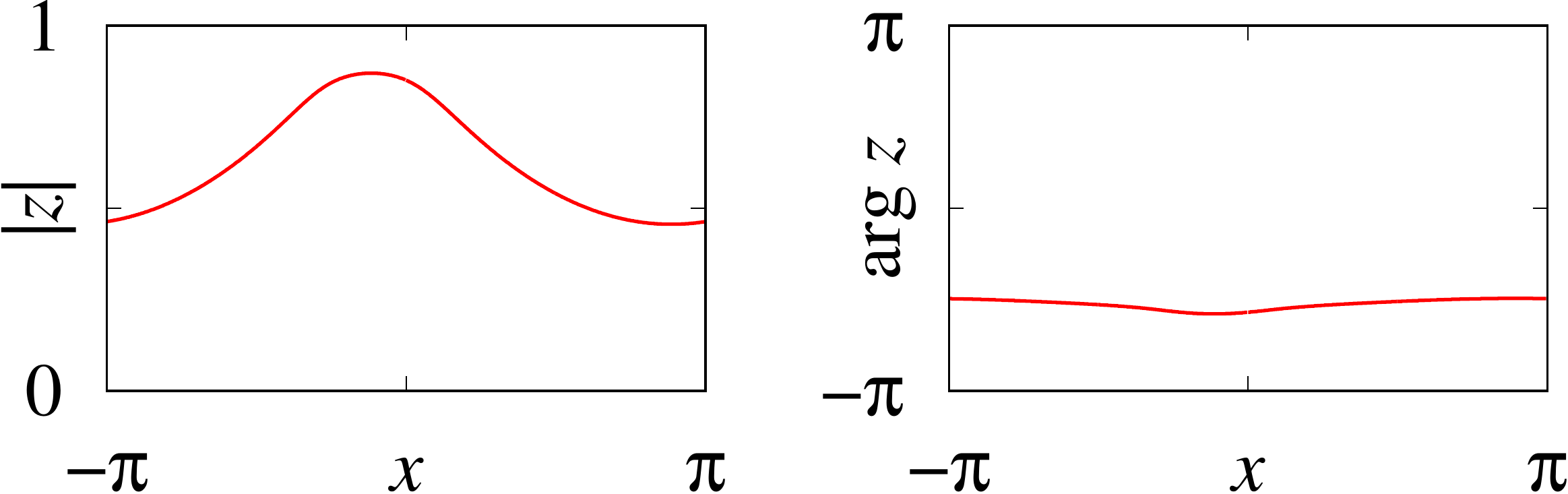}
\end{minipage}
\caption{Speed ($s$) and complex-phase velocity ($\Omega$) of the travelling wave for $\gamma = 0.02$.
Other notations the same as in Fig.~\ref{Fig:Diagram:0_01}.}
\label{Fig:Diagram:0_02}
\end{figure}

\begin{figure}[ht!]
\begin{minipage}[c]{0.26\textwidth}
\includegraphics[height=\textwidth,angle=270]{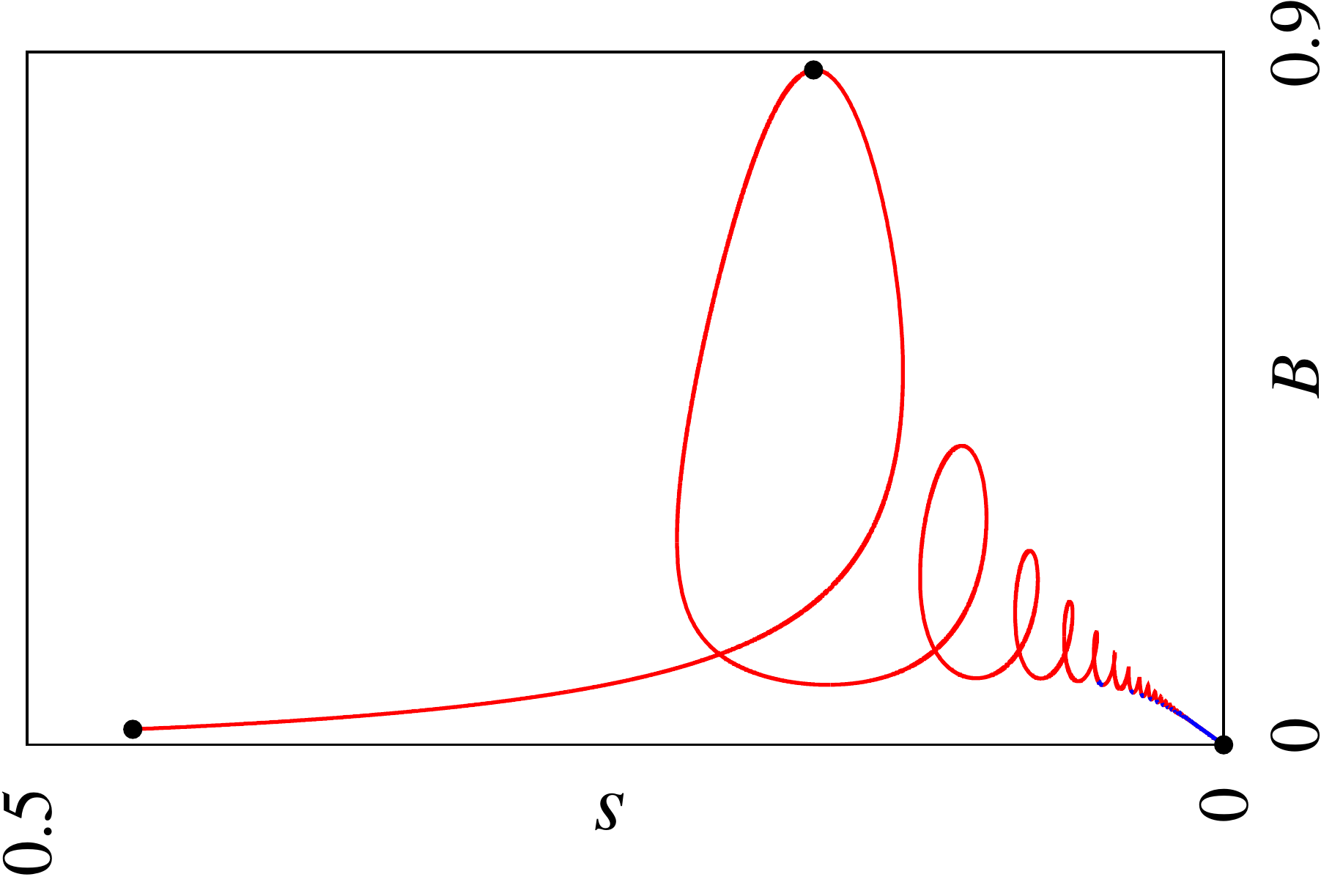}
\end{minipage}
\hspace{0.02\textwidth}
\begin{minipage}[c]{0.26\textwidth}
\includegraphics[height=\textwidth,angle=270]{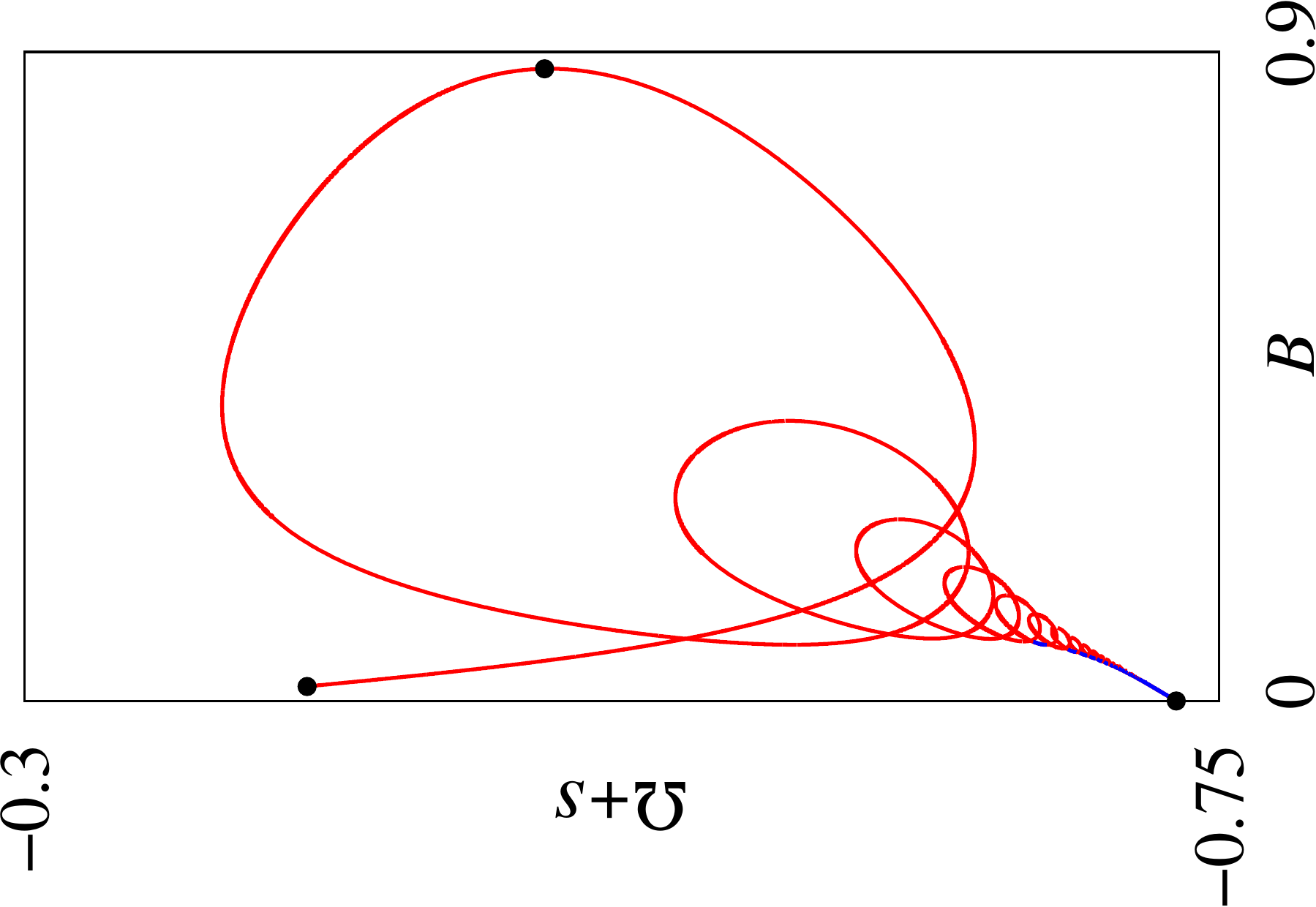}
\end{minipage}
\hspace{0.02\textwidth}
\begin{minipage}[c]{0.4\textwidth}
\includegraphics[height=\textwidth,angle=270]{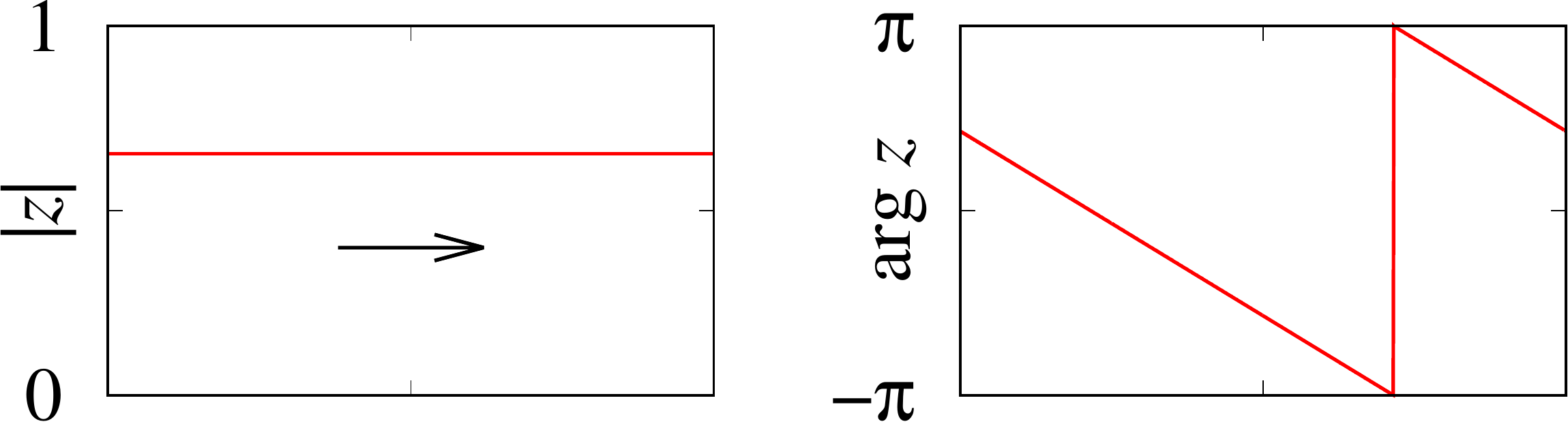}\\[2mm]
\includegraphics[height=\textwidth,angle=270]{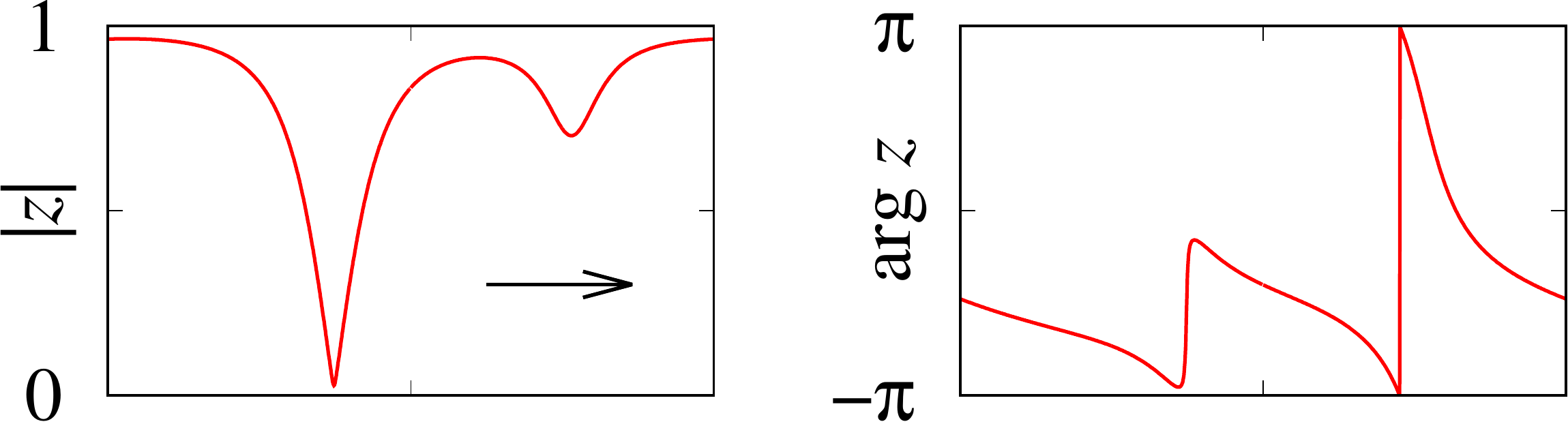}\\[2mm]
\includegraphics[height=1.01\textwidth,angle=270]{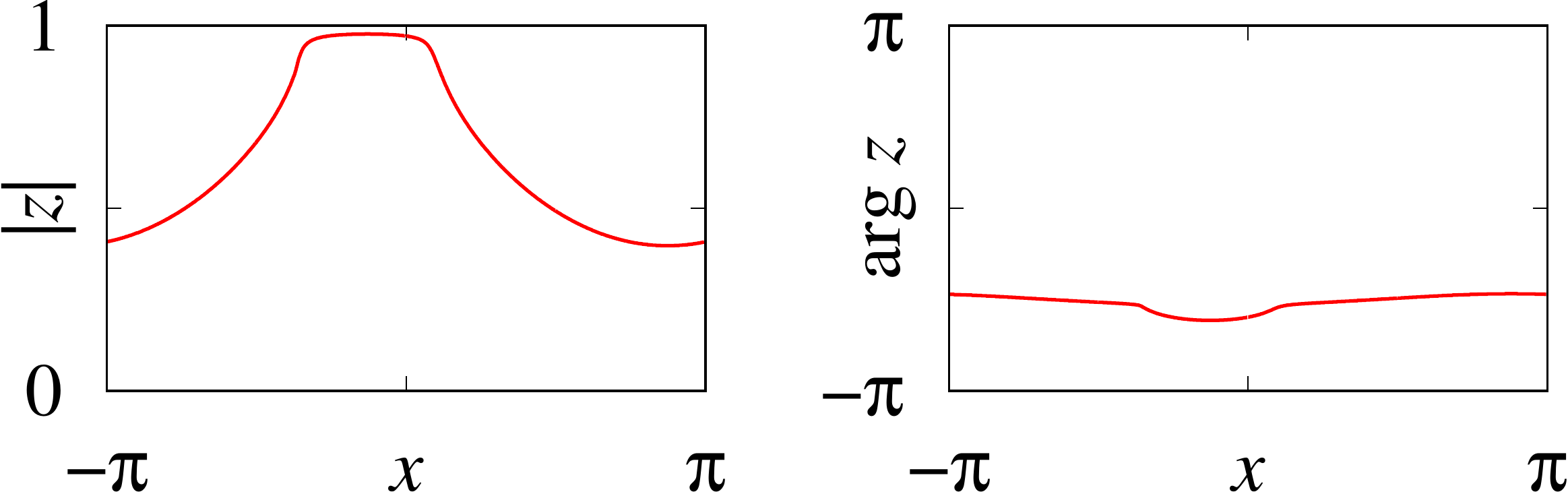}
\end{minipage}
\caption{Speed ($s$) and complex-phase velocity ($\Omega$) of the travelling wave for $\gamma = 0.005$.
Other notations the same as in Fig.~\ref{Fig:Diagram:0_01}.}
\label{Fig:Diagram:0_005}
\end{figure}

The situation that the branch of travelling waves occurs as a ``bridge'' between a standing wave ($s = 0$)
and some splay state is also observed for larger and for smaller values of~$\gamma$,
see Figs.~\ref{Fig:Diagram:0_02} and~\ref{Fig:Diagram:0_005}.
However, the smaller is $\gamma$ the more oscillatory is the corresponding $s$-versus-$B$ diagram.
Moreover, the maximal twist found on the branch of travelling waves typically grows for decreasing $\gamma$.
For example, for $\gamma = 0.02$ we find only twist-$0$, twist-$1$ and twist-$2$ waves.
In contrast, for $\gamma = 0.005$ we find travelling waves with twists up to $5$.
%%%%%%%%%%%%%%%%%%%%%%%%%%%%%%%%%%%%%%%
%%%%%%%%%%%%%%%%%%%%%%%%%%%%%%%%%%%%%%%
%
%%%%%%%%%%%%%%%%%%%%%%%%%%%%%%%%%%%%%%%
%%%%%%%%%%%%%%%%%%%%%%%%%%%%%%%%%%%%%%%

\section{Discussion}
\label{Sec:Discussion}

Let us summarize the main results of this paper.
In Section~\ref{Sec:Riccati} we have outlined a wide class of periodic complex Riccati equations,
whose Poincar{\'e} maps are represented by hyperbolic or loxodromic M{\"o}bius transformations,
which map the closed unit disc $\overline{\mathbb{D}}$ into itself.
We have shown that this class includes several types of the Ott-Antonsen equation
used in the analysis of the dynamics of phase oscillator networks and networks of theta neurons.
In Section~\ref{Sec:Moebius} we explained how the properties of M{\"o}bius transformations
can be exploited to calculate periodic solutions of the complex Riccati equation
by solving at most four intial value problems for this equation.
Finally, a practical application of the latter fact was demonstrated in Section~\ref{Sec:Chimera},
where we derived the self-consistency equation for travelling chimera states
arising in a ring network of phase oscillators with asymmetric nonlocal coupling,
and used it to calculate several complete branches of such states.

It is likely that the semi-analytical method of this paper can be generalized
to study more complex coherence-incoherence patterns in large networks of phase oscillators,
including breathing, pulsating and alternating chimera states~\cite{Lai2009,Lai2009a,Ome2020a},
as well as moving chimera states on two- and three-dimensional oscillator lattices~\cite{Bat-GCO2021,MaiSOM2015,LauD2016}.
Another class of potential applications is concerned with Proposition~\ref{Proposition:Sln_},
which can be useful in the study of moving and oscillatory bump states in theta neuron networks~\cite{LaiO2020,Lai2015}.
Moreover, recalling one of the recent applications of the Ott-Antonsen method
to networks of quadratic integrate-and-fire neurons~\cite{MonPR2015,ByrAC2019,SchA2020},
we expect that our method can also be adapted for such systems as well.
We plan to report on these issues in future work.

\section*{Acknowledgments}

This work was supported by the Deutsche Forschungsgemeinschaft under Grant OM 99/2-2.

%\cite{AbrMSW2008,SudO2018}.

\end{document}